\documentclass{bmcart}

\usepackage[utf8]{inputenc} 
\usepackage{dsfont}
\usepackage[pdftex]{graphicx}
\usepackage{tabularx}
\usepackage{color}
\usepackage{amsmath} 
\usepackage{amssymb}
\usepackage{amsxtra}
\usepackage{setspace}
\usepackage{rotating}
\usepackage{algorithm}
\usepackage{algorithmic}
\usepackage{booktabs}
\usepackage{cite,multirow}
\usepackage{amsthm}
\usepackage{setspace}

\newtheoremstyle{mystyle}
{3pt}
{3pt}
{\it}
{}
{\bf}
{:}
{.5em}
{}
\theoremstyle{mystyle}

\newtheorem{problem}{Problem}

\startlocaldefs
\endlocaldefs

\begin{document}

\begin{frontmatter}

\begin{fmbox}
\dochead{Research}


\title{{QoE} Optimization of Video Multicast with Heterogeneous Channels and Playback Requirements}


\author[
   addressref={aff1},                   
   corref={aff1},                       
   email={ali.bakhshali@queensu.ca}   
]{\inits{AB}\fnm{Ali} \snm{Bakhshali}}
\author[
   addressref={aff1},
   email={}
]{\inits{WYC}\fnm{Wai-Yip} \snm{Chan}}
\author[
   addressref={aff1},
   email={}
]{\inits{SDB}\fnm{Steven D} \snm{Blostein}}
\author[
   addressref={aff2},
   email={}
]{\inits{YC}\fnm{Yu} \snm{Cao}}


\address[id=aff1]{
  \orgname{Department of Electrical and Computer Engineering, Queen's
University}, 
  \postcode{ K7L 3N6}                                
  \city{Kingston, ON},                              
  \cny{Canada}                                    
}
\address[id=aff2]{%
  \orgname{Huawei Technologies},
  \postcode{ K7L 3N6}                                
  \city{Kanata, ON},                              
  \cny{Canada}          
}


\begin{artnotes}
\end{artnotes}

\end{fmbox}


\begin{abstractbox}

\begin{abstract} 
We propose an application-layer forward error correction (AL-FEC) code  rate allocation scheme to maximize the quality of experience (QoE) of a video multicast. The allocation dynamically assigns multicast clients to the quality layers of a scalable video bitstream, based on their heterogeneous channel qualities and video playback capabilities. Normalized mean opinion score (NMOS) is employed to value the client's quality of experience across various possible adaptations of a multilayer video, coded using mixed spatial-temporal-amplitude scalability. The scheme provides assurance of reception of the video layers using fountain coding and effectively allocates coding rates across the layers to maximize a multicast utility measure. An advantageous feature of the  proposed  scheme is that the complexity of the optimization is independent of the number of clients. Additionally, a convex formulation is proposed that attains close to the best performance and offers a reliable alternative when further reduction in computational complexity is desired. The optimization  is extended to perform suppression of QoE fluctuations for clients with marginal channel qualities. The scheme offers a means to trade-off service utility for the entire multicast group and clients with the worst channels. According to the simulation results, the  proposed optimization framework is robust against source rate variations and limited amount of client feedback. 
 
%
%
\end{abstract}




\end{abstractbox}
%

\end{frontmatter}




\section{Introduction}

\subsection{Motivation}
Multimedia delivery systems can be optimized to maximize the overall throughput (\emph{best-effort})  or to satisfy client quality of experience (QOE) demands (\emph{QoS-guaranteed}). QoE-guaranteed optimizations may suffer from being overly constrained,  especially in large-scale multicasts.  Tracking the media processing capability, QoE demand, and channel quality of every client can be daunting, prompting the search for better trade-offs  between bandwidth usage efficiency and optimization complexity. Sometimes no feasible solution  exists due to bandwidth limitations and/or clients with poor channels that require forward error correction (FEC) codes with exceedingly large overheads. Therefore, having a screening process to reduce excessive QoE demands is essential, especially in large-scale multicasts. %
One may utilize a mechanism to dynamically assign clients to available media quality levels in order to improve resource utilization efficiency. For example, using scalable bit streams, the multicast server may drop the highest enhancement layers when relatively few users with high quality channels and high resolution displays exist. The saved transmission resources could be redeployed to serve  clients with poor  channels. The multicast optimization needs to be performed repeatedly due to client channel and source bitstream variations, as well as  to account for clients dynamically joining or leaving the multicast at random times. Thus, low complexity optimization methods are required. %

In point-to-multipoint services such as multicast, the transmission to the multicast clients may traverse different paths. As a result, the end-to-end transmission channels may exhibit diverse behaviors and capacities.  End-to-end QoE can be assured by providing sufficient error protection.  Feedback based error correction such as automatic repeat request (ARQ) and Hybrid-ARQ \cite{HARQ} may not be feasible due to latency and possible feedback implosion at the multicast server.  An alternative which avoids these problems is to employ FEC coding. In multicasting, we are faced with an ensemble of channels with different loss processes and require FEC that is ``universally'' efficient.  Fortunately, fountain codes \cite{Fountain_Codes} have been demonstrated to well approximate the ideal.  With fountain codes, the receiver can recover the source symbols with high probability when the number of correctly received code symbols is slightly larger than the number of source symbols.  Crucially, this recovery capability is independent of the loss pattern or channel memory. One implication of this ``independence'' from channel memory is that clients connected to distinct channels with differing memory behaviors that inflict the same amount of loss will see the same throughput. 

This paper is concerned with efficient application of fountain codes as an application-layer FEC (AL-FEC) code to meet the QoE demands of video multicast clients with heterogeneous channels and video quality requirements. 
 This approach offers the following advantages :  %
1) service versatility since the service is agnostic to the underlying network infrastructures, enabling  clients to join the multicast through a variety of network connections; %
2) quick service deployment or reconfiguration, eliminating the wait for infrastructure upgrade and enabling quick launch of third-party services; and 3) extending  the capability of an existing network (infrastructure) \cite{Luby_DVBT_2009}.

\subsection{Related Approaches}

Multicast schemes have evolved with advances in source and channel coding techniques.  
Receiver-driven layered multicast (RLM) \cite{RLM} is a landmark technique for multicasting to clients with heterogeneous  channels. RLM is a ``client pulled'' scheme suitable for large-scale multicast over the Internet. Subsequently, unequal error protection (UEP) was proposed \cite{UEP_Mohr} and its application to multimedia transmission was studied \cite{UEP_Mohr_LayeredVideo_TMM2001}. Further works largely fall into one of the following three categories: AL-FEC design for UEP \cite{Sachs_MDC,SVM_EWF, SlidingWF_codes, UEP_Fountain_Ahmad_TMM11,UEP_Raptor_OP_2013,Wang_Rateless_UEP_ISM11,Wagner_MultiServer_2006},  link
layer scheduling \cite{Wu_OptLayerIPTVWIMAX_TMM11,Xu_ScVidMulBroadChan_GlOBECOM09,Kue_Utility2008, Sharangi_Wimax2011,Vukadinovic_Multi_SVC}, and  joint source-channel coding \cite{Ji_Joint_TMM12}. In practice, system design and provisioning usually prefer separate source and channel coding as well as low computation complexity.  

Fountain codes are employed in many current multimedia delivery standards \cite{DVBH,3GPP_stand}  due to their structural benefits, e.g.,  linear time encoding/decoding algorithms and small overhead \cite{Raptor_linear_Time,Luby_Pe}.
 Digital fountain based approaches in the AL-FEC design category  \cite{SVM_EWF, SlidingWF_codes,UEP_Fountain_Ahmad_TMM11} mainly rely on altering the degree distribution and source symbol selection process, to provide UEP across different source layers.  %
 In \cite{Yen_LT_ModDist_UEP_TMM13} the fountain code degree distribution is optimized to provide short code length performance. 
The advantage of using rateless codes over conventional Reed-Solomon codes in providing graceful-degradation was reported in  \cite{Wang_Rateless_UEP_ISM11}. In \cite{Wagner_MultiServer_2006} UEP and rateless coding are utilized in streaming a scalable video from multiple servers. 
 This work aims to  maximize the probability of successful decoding through proper rate allocation amongst video layers of different servers. 
 Note that none of the above fountain code based works consider client channel heterogeneity in their design.  Moreover, these schemes treat only one scalability dimension (PSNR) and do not optimize the visual perceptual quality.

There are a number of  notable link-layer scheduling algorithms for multimedia multicast.
 A best-effort optimization framework is proposed in \cite{Wu_OptLayerIPTVWIMAX_TMM11} for Internet protocol television broadcast over WiMax channels with consideration of capacity variation in the multicast channel. %
Sharangi {\it et al.} \cite{Sharangi_Wimax2011} proposed a scalable video transmission scheduling optimization scheme for multiple multicasts to share a set of WiMax timeslots such that  the average utility of the multicasts is maximized. 
A similar work with a more elaborate model of physical layer parameters and channel effects is proposed by  Vukadinovic {\it et al.} \cite{Vukadinovic_Multi_SVC}. 
  While our problem (described below) and \cite{Sharangi_Wimax2011,Vukadinovic_Multi_SVC} both strive to balance  serving individual clients versus overall throughput, for our problem  the individuals are clients with heterogeneous channels and playback requirements within a multicast, whereas for \cite{Sharangi_Wimax2011,Vukadinovic_Multi_SVC}  the individuals are distinct multicasts each of which targeting  one channel and one media quality. %

Several multicast schemes benefiting from application-layer FEC and  file delivery over unidirectional transport (FLUTE)\cite{FLUTE_2012} have been recently introduced\cite{deFez_Adaptive_Download_Wireless_TMM14,Multicast_eMBMS_LTEa_12}. Adoption of dynamic adaptive streaming over HTTP (DASH) to support  multicast services is discussed in \cite{DASH_Stockhammer_VCIP12,MPEG_DASH_WIFI_ICCT13}. A hybrid  multicast architecture based on FLUTE and DASH is proposed in \cite{Hybrid_FLUTE_DASH_2012} where FLUTE provides multicasting with application-layer FEC and DASH is utilized for retransmission of lost frames over a unicast channel. Bouras {\it et al.} \cite{Bouras_AL_FEC_Multicast_LTE_13} experimentally assessed the efficacy of using standard raptor\cite{Raptor_Codes} codes as application-layer FEC codes for multicasting video over 3GPP LTE (Long Term Evolution) wireless networks. 
The assessment employs non-scalable low-bit-rate video and no service optimization is performed. 

\subsection{Proposed Approach}

In this paper, we formulate and solve an application-layer fountain code rate allocation optimization problem for multicasting a scalable coded video (SVC) stream with the aim to maximize service utility. We consider client heterogeneity in terms of channel quality diversity and media decoding capability. Application layer  multicast obviates the need to access the lower network layers in order to control the transmission scheme. Clients may be connected to the service using different physical channels. For instance, mobile clients may be able to access multiple network infrastructures and  engage in ``vertical handoffs'' across different networks. From the perspective of the multicast service, the end-to-end path to individual clients may traverse different network infrastructures with their underlying physical-layer error protection mechanisms. For the purpose of our AL-FEC coding optimization, the net effect of the end-to-end channel capacity is parameterized in the form of a ``reception coefficient'' (RC). The RC parameter enables the application layer to use a memoryless erasure channel model (see (10) below) to represent, for instance, lower layer FEC decoding performance in cellular networks or packet losses on the Internet. The diversity of client channel capacities is modeled using probability distributions.  The utility is based on using an objective video quality measure to value  client satisfaction across different possible adaptations of the video layers. A client may have a specific playback profile, which could be elastic in the sense that the client may be willing to accept (or even reject) playback of various layer adaptations, with corresponding degrees of utility gained. 
 The allocation is performed to maximize a utility measure that permits balancing between individual client utility and serving as many clients as possible.  Our problem provides an
answer to the question: given an application-layer multicast service bandwidth, a population of clients with heterogeneous end-to-end channels and devices (with different video playback capabilities), determine how best to provision fountain codes across the video layers in order to serve as many clients as possible while
meeting their video perceptual-quality demands. A byproduct of our problem solution is indicating which
clients cannot be served to meet their desired viewing quality.  

Our problem is fashioned to enable using standard fountain codes or their equivalent. We believe this is a more attractive proposition for multicast equipment/service engineering than using customized fountain codes. A client utility measure is defined   based on a visual perceptual model \cite{NMOS_Xue1,NMOS_OU2} that admits mixed spatial-temporal-amplitude scalability. Our multicasting framework also offers the flexibility to admit other 
advanced video quality assessment models for mixed-scalability video. An advantageous feature of the proposed method is that the optimization complexity does not increase with the number of clients, a property particularly appealing for  large-scale multicasts. Moreover, by employing statistical modeling of client reception capabilities, the optimization can be performed with different resolutions to trade-off  complexity and performance. 
The reliability of decoding the video layers in terms of outage probability (OP) is enforced to be commensurate with the probabilistic decoding nature of rateless codes. 
Compared to the previous multicast optimization techniques based on fountain codes in \cite{SW_Raptor_SVC_UEP_TIP2010, SVM_EWF,UEP_Fountain_Ahmad_TMM11}, our work considers clients with heterogeneous channels and  video-playback quality demands,  and benefits from a simple yet accurate  model \cite{Bakhshali_QBSC12} of the client decoding outage probability. The QoE of the proposed multicast scheme has both guaranteed and best-effort aspects. The qualities of the different video layers are guaranteed, provided the client's channel has commensurate capacities. The best layer the client can access also depends on the client population channel qualities and demand profiles. Another aspect of our framework is that it does not require altering the video bit stream or rateless code, avoiding compatibility issues with existing and future standards, e.g., \cite{HEVC,RaptorQ,Spinal_Codes}.

Additionally, we extend our previous work on video multicast optimization  \cite{Bakhshali_MMSP12} to suppress temporal quality fluctuations caused by source bit rate variation.  
By utilizing a quality-aware optimization that admits source scalability, the proposed scheme provides a range of trade-offs between transmission resource utilization  efficiency and stable client video playback quality.  
With some simplifications, we obtain a convex optimization problem. It turns out that the solution of the convex problem is a highly accurate approximation. %

The rest of this paper is organized as follows. Section \ref{Section_Setup} is devoted to the general problem formulation as  well as a convex formulation that admits lower computation with  moderate loss in accuracy. In Section~\ref{Section_CliDis}  we extend our formulation to a dynamic optimization that considers client dissatisfaction due to video quality fluctuations. In Section~\ref{Section_Util} we assign values to the client utility parameters in our formulated problem using a recently developed video quality metric. The performance of the  proposed optimization framework is evaluated in Section~\ref{Section_USim}. Finally, conclusions are drawn in Section~\ref{Section_Conc}. The basic notations used in this paper are listed
in Table~\ref{notations}.

\section{Proposed Multimedia Multicast with Heterogeneous Clients}
\label{Section_Setup}
\subsection{System Setup}
Fig. \ref{fig:Scheme} illustrates  the system setup. A media server is responsible to provide various terminal (user device) classes with a multilayer media, e.g., an H.264/SVC encoded video stream. A hybrid network of wired and wireless clients with heterogeneous channels is depicted. For  encoding, a sequence of video frames is partitioned into consecutive time segments. Each segment, which may comprise the frames say over a one-second interval, is encoded into a scalable bitstream. The generated bitstream embeds $L$ layers with $S_l$ source symbols per layer $l$, $l=1,...,L$. While the base-layer  is essential, the enhancement layers introduce  higher spatial or temporal resolution, or  finer  quantization resolution without altering the spatio-temporal resolution of the preceding layer. We assume that successful decoding of any layer relies on successful decoding of all of its preceding layers. This implies that layers with lower indices are more important in the decoding process.
Fountain coding \cite{Fountain_Codes} in the form of raptor codes is applied to every layer of the bitstream to provide protection against erasures caused by channel errors in the physical layer.  The code for layer $l$ receives $S_l$ source symbols and  generates $N_l$ encoded symbols. Unlike conventional Reed-Solomon codes, fountain codes can potentially  generate an infinitely large code sequence, making the code rate $S_l/N_l$ elastic, or the code ``rateless''. Generation of the rateless code sequence is determined by specifying a degree distribution  and a random number generator. Here, we exploit the elastic property by choosing the code rate $S_l/N_l$ to best suit an optimization objective.  
 Standardized raptor codes \cite{Fountain_OV}  have been optimized so that a receiver that correctly receives $K_l=S_l(1+\epsilon)$ encoded symbols from the transmission can recover the message, with $\epsilon>0$ representing a small overhead typically below $2\%$. 
  Successful decoding is probabilistically ensured by the total number of transmitted symbols successfully recovered by the receiver \cite{Bakhshali_QBSC12}. For practical considerations, we assume that $N_l$  encoded symbols are transmitted for the $l^{\text{th}}$ layer such that $\sum_{l=1}^LN_l\leq N_{\text{\text{max}}}$.  $N_{\text{\text{max}}}$, which we call the ``service bandwidth'', is set as part of the service provisioning and may depend on the bandwidth available to the server, the temporal duration of the video segment, and other factors. For example, consider a video sequence which is partitioned into segments each with  $T_{\text{seg}}$ second duration, and a server allocated bandwidth of $\Omega$ bit/s. Assuming that each symbol comprises $B$ bits, the maximum number of  available transmission symbols for each video segment is
\begin{align}
N_{\text{\text{max}}}=\left\lfloor \frac{\Omega \, T_{\text{seg}}}{B}\right\rfloor
\label{Nmax_BW}
\end{align}
and can be chosen and even varied across segments to meet deadline requirements in streaming applications.
The multicast clients are modeled by $M$  classes of media players, each class comprising players that are capable of decoding the media up to layer $h_m\in\{1,...,L\}$, $m=1,\dots,M$, and have commensurate display resolutions. Classes are indexed in increasing order $h_1<h_2,...,<h_M$. %
 Clients with high definition (HD) displays may demand decoding up to a HD layer,  while  smart-screen and portable device  users may demand standard definition (SD) or a lower resolution to suit their application memory capacity and/or power consumption policies. %
For example in Fig.~\ref{fig:Scheme}, multicast transmission of a source with $L=8$ layers to $M=3$ classes of users is considered. Mobile and portable TV clients can potentially decode the video up to layers $h_1=3$ and $h_2=6$, respectively, while all 8 layers are decodable by HD clients ($h_3=L=8$). 
Clients may also have different  reception capabilities,
e.g., due to having different bandwidths, antenna systems, and radio propagation characteristics.
A reception coefficient (RC) $0 \leq \delta \leq 1$ is used to model the client reception capability, where $1-\delta$ is the application-layer packet loss rate due to  loss phenomena in the lower layers. %
We assume memoryless erasure channels (MECs) with  independent and identically distributed (i.i.d.) erasures between the server and the clients.  %
A client channel with RC $\delta_c$ has an erasure rate of $1-\delta_c$ and receives an {\it expected} number of $\delta_c N_{\text{\text{max}}}$ transmitted fountain symbols in a transmission period
of one video segment. Note that the actual number of the correctly received symbols depends on
the channel symbol erasure events. 
We  define the cumulative distribution function (CDF) of the channel quality of  class $m$ clients as $F_{m}(\delta), m=1,\dots,M$. Additionally, prior class probabilities $\pi_m>0$, $m=1,...,M$  with $\sum_{m=1}^M \pi_m=1$ are used to reflect the distribution of client population across different classes.

The media layers are not of the same importance to the clients. %
QoE for a client depends on the probability of successfully acquiring the layers the client desires. It is possible for one or more desired layers not to be served due to resource or channel limitations. For those clients that are served a particular layer $l$, the probability of failing to decode the layer can be limited by setting  outage probability constraints $P^l_{\text{out}}, 1 \leq l \leq L$
.  While it is conceivable that the clients desiring the same layer might want different levels of decoding assurance, for simplicity we assign one assurance level, in the form of probability $1-P^l_{\text{out}}$, to each media layer. %
Ideally, every additional encoded symbol drawn from a digital fountain improves the decoding probability of the code. Thus, if $N_{\text{\text{max}}}$ is allowed to be sufficiently large, all clients with non-zero RC will eventually achieve the targeted QoS. However, in a more realistic scenario with finite transmission resources  $N_{\text{max}}$, and any given set of $N_l, l=1,...,L$ with $\sum_{l=1}^L N_l=N_{\text{\text{max}}}$, we can find a set of  minimum needed reception coefficients (MNRCs) $\delta_{l}$ such that those clients with RC $\delta_c < \delta_{l}$ and desiring the layer $l$ media  will not reach the layer-decoding assurance probability $1-P^l_{\text{out}}$. 
Since successful decoding of all layers $j=1,...,l$, is necessary in order to enjoy the media quality of layer $l$, we impose an unequal error protection (UEP) condition\begin{align} \label{UEP_CH5}
0 <\delta_{1} \leq \delta_{2} \leq \ldots \leq  \delta_{L} \leq 1.
\end{align}
Later, we prove that this condition is necessary for optimal  utilization of transmission resources while simplifying  the utility function.

\subsection{Utility Function}
\label{Section_BestEff}

Let $u_{m,l}$ be the utility for  class $m$ clients decoding layer $l$ with decoding failure probability guaranteed to be below a given outage probability threshold. %
 Our ``utility'' differs from the conventional average utility found in best effort QoE formulations, wherein utilities associated with unacceptable decoding failure probabilities are included in the utility averaging. $u_{m,l}$ is a function of the number of clients who are able to decode layer $l$ under the guarantee, as well as the amount of utility they  gain,
\begin{align}
u_{m,l}&= \alpha_{m,l}\int_{0}^1f_m(\xi) \mathcal{I}(\prod_{j=1}^l{\big[1- P(S_j,N_j,\xi) \big] \geq (1-P_{\text{out}}^l)}) \quad d\xi.  \label{equ:Servj}
\end{align}
Here, $f_m(\delta)$ is the RC probability distribution of clients in class $m$, $\mathcal{I}(.)$ is the indicator function,  $P(S_j,N_j,\delta)$ is the probability of failing to decode the fountain code in layer $j$,  with $S_j$ source symbols and $N_j$ transmitted symbols, for a client with RC $\delta$, and $\alpha_{m,l}$ is the incremental utility gained by a class $m$ client  after decoding layer $l$, provided that all preceding  layers are successfully decoded. 
$\alpha_{m,l}$ is obtained from the utility-rate function of each client class, $\mathcal{U}_m(R_l)$, i.e.,
\begin{equation} \label{alpha_u}
\alpha_{m,l}=\mathcal{U}_m(R_l)-\mathcal{U}_m(R_{l-1}) ,\quad \forall \ \ l,m>0.
\end{equation}
We show in Section \ref{Section_Util} a specific way of using this function to optimize viewing experience. In \eqref{alpha_u}, 
$R_l=\sum_{k=1}^lS_k/T_{\text{seg}}$ is  the cumulative source symbol  rate up to layer $l$  with $R_0\triangleq0$ and $\mathcal{U}_m(0)\triangleq0$, $\forall m$. 
The product term within the indicator function in \eqref{equ:Servj} provides the probability of successfully decoding  all layers up to and including layer $l$.  %
 With the MNRCs $\delta_l$ defined earlier, we can write $\prod_{j=1}^l{\big[1- P(S_j,N_j,\delta_l) \big] = (1-P_{\text{out}}^l)}$  and then rewrite \eqref{equ:Servj} as
\begin{align} \nonumber
u_{m,l}&= \alpha_{m,l}\int_{0}^1f_m(\xi) \mathcal{I}(\xi \geq \delta_l)\, d\xi\\
&=\alpha_{m,l}\int_{\delta_l}^1f_m(\xi)\ d\xi=\alpha_{m,l} [1-F_{m}(\delta_{l})]. \label{eq4}
\end{align}
We obtain the utility of class $m$ clients $U_{m}$ by accumulating the guaranteed utility of all useful layers. However, we should make sure that the incremental utilities $\alpha_{m,l}$ for enhancement layer $l$ contributes to $U_m$ only when the clients can reliably decode the preceding layers. The UEP conditions embodied in \eqref{UEP_CH5} represent the hierarchical decoding dependencies of the scalable video layers and provide the needed assurance. 
\begin{align}
U_{m}&=\sum_{l=1}^{h_m} u_{m,l}=\sum_{l=1}^{h_m} \alpha_{m,l} [1-F_{m}(\delta_{l})]. \label{Um}
\end{align}
Not all the video layers can be useful for the clients of a class due to screen resolution or other playback constraints. Therefore, in \eqref{Um}, $h_m\leq L$ denotes the highest video layer which can contribute to the utility of class $m$ clients. 
%

Finally, the overall utility is obtained by summing over the utilities of all client classes using the prior class probabilities $\pi_m>0$, $m=1,...,M$,
\begin{align} \nonumber
{U}_{\text{total}}&=\sum_{m=1}^M \pi_m U_m=\sum_{m=1}^M \pi_m \sum_{l=1}^{h_m} \alpha_{m,l}[1- F_{m}(\delta_{l}) ] \hspace{-1cm} \ \\[-.1cm]
&\equiv 
{U_{\text{max}}}- \sum_{m=1}^M \sum_{l=1}^{h_m} \hat{\alpha}_{m,l} F_{m}(\delta_{l})   \label{equ:ServTot}
\end{align}
where $\pi_m$ is absorbed into $\alpha_{m,l}$ by defining $\hat{\alpha}_{m,l}=\pi_m \alpha_{m,l}$ and ${U_{\text{max}}}=\sum_{m=1}^M \sum_{l=1}^{h_m} \hat{\alpha}_{m,l}$. Note that ${U_{\text{max}}}$ is an upper bound on the deliverable utility, and only depends on the media source  and the priors. This bound is achievable if the MNRCs $\delta_{h_m}, m=1,...,M$, are small enough so that no client has to settle for a quality layer lower than their maximum desired quality. However, this may not be possible since the service bandwidth $N_{\text{max}}$ and the OP constraints prevent the MNRCs from becoming arbitrarily small. As a result, clients with poor RCs may end up being not served  their most desired video quality, or even worse, being unable to decode the base layer. ${U}_{\text{total}}$ is to be maximized, as shown below. We  emphasize that the problem at hand  is efficient utilization of the multicast service bandwidth $N_{\text{max}}$ to provide guaranteed utility to individual multicast clients while serving as many clients as possible. However, for a given $N_{\text{max}}$ and set of client RC distributions, the problem solution  may not be able to service a portion of the clients with exceedingly poor channels. These clients may be served by increasing $N_{\text{max}}$ or providing alternate solutions, e.g., unicast (re)transmission, peer-assisted repair \cite{IPTV_Mlticast_peer_assisted_CSVT12}. Such solutions are outside the scope of this paper.

\subsection{Outage Probability}
Let $P(S,N,\delta)$ be the probability that a client fails to decode the $S$ information symbols,  given the client's RC $\delta$ and the number of transmitted symbols $N$. 
 The performance of a rateless decoder in decoding a  source with $S$ information symbols after receiving  $K$ code symbols is given by the decoding failure probability function  $P_f(S,K)$. Assuming interleaving is used if needed, we consider a memoryless erasure channel (MEC) with symbol erasure rate $1-\delta\in[0 \;, \;1]$ assumed to be fixed during the transmission period of a video segment. For a given number of transmitted code symbols, the outage probability can be obtained from 
 \begin{align} \label{ExpInt}
P(S,N,\delta)=\mathbb{E}_{K|N} [P_f(S,K)]
\end{align}
with erasure probability $1-\delta$ and i.i.d. erasure events, $K$ is a binomial random variable. Moreover, the decoding failure probability of rateless codes can be modeled by \cite{Luby_Pe}
\begin{equation}\label{Erasure}
P_f(S,K)=\begin{cases}
1, & \mbox{if } K\leq {S}, \\
a b^{K-{S}}. & \mbox{if } K>{S},
\end{cases} \hspace{.5cm} \vspace{-.1cm}
\end{equation}
where $a>0$ and $0<b<1$ vary with the rateless code structure, particularly the degree distribution, and the precode rate. For example, $a=0.85$ and  $b=0.567$ were used for the raptor code in~\cite{Luby_Pe}. Combining  \eqref{ExpInt} with \eqref{Erasure}, we obtain the outage probability of a rateless coded source over a MEC
\begin{align} \nonumber
P(S,N,\delta)&=\sum_{k=0}^{N} {N \choose k}\delta^{k}(1-\delta)^{N-k} P_f(S,k)\\ 
&=\text{Bin}_{N,\delta}(S)+\sum_{k=S+1}^{N}{N\choose k}\delta^{k}(1-\delta)^{N-k}ab^{k-S}. \label{Pout1}
\end{align}
Here, $\text{Bin}_{N,\delta}(.)$ is the binomial CDF with parameters $N$ and $\delta$. Despite its accuracy, the closed-form representation in \eqref{Pout1} is not convenient for optimization in which one needs to express other parameters as an explicit function of the OP. To deal with this shortcoming, the following parametric model that was previously derived in \cite{Bakhshali_QBSC12 } offers a convenient approximation of \eqref{Pout1}:
\begin{align} \label{OutMod}
\widetilde{P}(S,N,\delta)= 0.5\exp\big[-\frac{\delta (N-{S}/{\delta})^H}{S (1-\delta)}\big] \hspace{.2cm} \text{for\ } \  N\geq {S}/{\delta}.
\end{align}
Note that $H\approx1.8$ for the rateless codes used in \cite{Luby_Pe}. As shown in Fig. \ref{OutFig}, this model accurately estimates the outage probability \eqref{Pout1} for various channel parameters.

 In summary,  we aim to maximize the utility in \eqref{equ:ServTot} subject to the bandwidth  and the UEP constraints defined in Section II.A. The first term in \eqref{equ:ServTot} is not a function of the optimization variables, MNRCs $\delta_{l}$, $l=1,...,L$. Hence,  the utility maximization  can be transformed into the following utility loss minimization problem:
\begin{problem}{(General formulation)} \label{final_Problem}
\begin{align} \nonumber
\ & \displaystyle \min_{\{\delta_{l}\}_{l=1}^L}  \; \ \sum_{m=1}^M \sum_{l=1}^{h_m} \hat{\alpha}_{m,l} F_{m}(\delta_{l}) &\\ \nonumber   \textrm{\emph{subject to}} \\
 \ \text{UEP Constraints: } &   \delta_{1}\ge0, \\ \nonumber
														&		\delta_{l}-\delta_{l+1}\leq0, \qquad l=1,...,L-1,\\\nonumber
														&  \delta_{L}\leq1, \\													\nonumber
												\ \text{BW Constraint. : } & 
  \displaystyle \sum_{l=1}^{L} N_l \leq N_{\text{max}}.
\end{align}
\end{problem}
\noindent 
 %

 We first consider using exhaustive search to solve Problem 1. The set of all $\delta_l, l=1,...,L$ satisfying the UEP constraints forms an $L$-simplex in $L$ dimension. For exhaustive search, the simplex volume is discretized using a $L$ dimensional cubic lattice $\mathcal{L}$ with $|\mathcal{L}|$ points. For each point in $\mathcal{L}$, say $\delta_l$, $l=1,...,L$, we first obtain the required per layer transmission resources $N_j$ in a forward procedure using 
\begin{align} \label{Reverse_outage}
N_l=\begin{cases}
P^{-1} (S_1,\delta_1,P_{\text{out}}^l) & l=1,\\ 
P^{-1}\bigg(S_l,\delta_l,1-\frac{1-P_{\text{out}}^l}{\prod_{j=1}^{l-1} [1- P(S_j,N_j,\delta_l)] }  \bigg)  & l\geq 2,
\end{cases}
\end{align}
wherein $P^{-1} (S,\delta,p)$ is the inverse outage probability function which yields the required number of transmitted symbols $N$ as a function of the number of source symbols $S$, the reception coefficient $\delta$, and the designated outage probability constraint $p$. A convenient closed form expression of $P^{-1} (S,\delta,p)$ is obtained by rearranging the terms in the approximated OP model \eqref{OutMod}.
Having $N_j, j=1,...,L$ in hand,  the bandwidth constraint is checked. If the constraint is satisfied, the  cost function is calculated; otherwise, the cost is set to infinity. For a sufficiently fine discretization, we regard the minimum cost point in $\mathcal{L}$ as the ``optimal'' solution. Note that by using the bandwidth constraint in the above manner, the exhaustive search can be conducted over $L-1$ dimensions. The complexity  $\mathcal{O}(D^{L-1})$ can be large, where $D$ is the number of grid points on each dimension.

 After  obtaining the  optimal  MNRCs, $\delta_l^*, l=1,..,L$, the corresponding transmission resources per layer $N_l^*, \forall l$ are obtained. 
Clients whose highest media quality demand is layer $l$ but whose RCs are below $\delta_l^*$ have to settle for the lower quality of layer $i$ where $i$ is the largest layer index with $\delta_i^*$ no greater than the client's RC. Ultimately, clients with RCs below $\delta_1^*$ are dropped from the multicast as they cannot  decode the base layer with the assured probability.

In contrast to other formulations such as \cite{Kue_Utility2008} in which clients are individually represented in the optimization, here multicast clients are grouped and represented by the distributions $F_{m}(\delta)$ and associated priors $\pi_m$. Consequently, the complexity of the proposed optimization is independent of the number of clients. Moreover, client-to-server feedback for the purpose of updating the RC distributions could be managed  without feedback implosion, e.g., the server could broadcast a threshold value and clients with a locally generated random number above the threshold would send  their RCs to the server.  This threshold is adapted to the multicast population size such that the server is not overwhelmed by excessive amount of feedback messages. The RC distributions  could be parametrized or discretized with a suitably chosen resolution to trade-off between computational complexity and accuracy. %

\subsection{Simplified Formulation} \label{Section_SimForm} 

Next, we exploit simplifications of the  outage probability constraints to obtain a problem formulation that is amenable to solution using gradient search. Let $Q_l(\delta)=\prod_{j=1}^l{\big[1- P(S_j,N_j,\delta) \big]}$ be the probability of receiving layers 1 to $l$. $Q_l(\delta)$ is monotonically non-increasing with $l$ and monotonically non-decreasing with $\delta$. Moreover, due to the fast-decaying nature of the decoding failure probability \eqref{Erasure}, $Q_l(\delta)$ exhibits an abrupt transition for $\delta$ in the neighborhood of  $\delta_l$. This can be seen from Fig.~\ref{SimpOutage_fig} which shows $Q_l(\delta)$ and $[1-P(S_l,N_l,\delta)]$ for a closely spaced set of  $\delta_l$'s. It can be seen from Fig.~\ref{SimpOutage_fig} that in the neighborhood of $\delta_l$, the factor $Q_{l-1}(\delta)=\prod_{j=1}^{l-1}{\big[1- P(S_j,N_j,\delta) \big]}$ is nearly one, and the transition behavior of $Q_l(\delta)$ is dominated by  $\big[1- P(S_l,N_l,\delta) \big]$. 
Hence, we can use the approximation
\begin{align}
\prod_{j=1}^l{\big[1- P(S_j,N_j,\delta_l) \big]} \approx 1- P(S_l,N_l,\delta_l). 
\end{align}
Consequently, for a given set of $\delta_l, l=1,...,L$, the per-layer transmission resources $N_j$ can be  obtained from 
\begin{align} \label{Reverse_outage2}
N_l=P^{-1}\bigg(S_l,\delta_l,P_{\text{out}}^l \bigg) , \quad l=1,...,L.
\end{align}
Using \eqref{OutMod} to estimate $N_l$ as a function of the outage probability we have
\begin{align}
    N_l &= S_l/{\delta_{l}}+\tau_{l}\sqrt[H]{\frac{1-\delta_{l}}{\delta_{l}}},   \label{OV_vs_delta_NC}
\end{align}
where 
\begin{align}
\tau_{l}=\sqrt[H]{-S_l \ln\left(2{P^l_{\text{out}}}\right)}, \quad P^l_{\text{out}}\leq0.5.
\end{align}
Using this the bandwidth constraint becomes
\begin{align}
   \sum_{l=1}^L \bigg(S_l/{\delta_{l}}+\tau_{l} \sqrt[H]{\frac{1-\delta_{l}}{\delta_{l}}} \bigg)\leq N_{\text{max}} , \quad 0<P^l_{\text{out}}\leq a. \label{N_vs_delta_simp}
\end{align}
As a result, a new optimization problem can be formulated.
\begin{problem}{(Simplified formulation)} \label{Simp_Problem}
\begin{align} \nonumber
\ & \displaystyle \min_{\{\delta_{l}\}_{l=1}^L}  \; \ \sum_{m=1}^M \sum_{l=1}^{h_m} \hat{\alpha}_{m,l} F_{m}(\delta_{l}) &\\ \nonumber   \textrm{\emph{subject to}} \\ \nonumber
 \ \text{UEP Constraints:\quad} &  \delta_{1}\ge0, \\ \nonumber
														&		\delta_{l}-\delta_{l+1}\leq0, \qquad l=1,...,L-1,\\\nonumber
														&  \delta_{L}\leq1, \\													\nonumber
                        %
												\ \text{BW Constraint:\quad} & 
  \displaystyle \sum_{l=1}^L \bigg(S_l/{\delta_{l}}+\tau_{l} \sqrt[H]{\frac{1-\delta_{l}}{\delta_{l}}} \bigg)\leq N_{\text{max}}.
\end{align}
\end{problem}
\noindent Unlike Problem 1, first order derivatives of the BW constraint can now be easily obtained. Hence, gradient descent algorithms with $\mathcal{O}(L \log(1/e))$ complexity, where $e$ is the required accuracy, can be deployed to solve Problem 2. Since Problem 2 may have multiple local minima, the quality of the gradient descent solution depends on the algorithm initialization. In the next section, a convex approximation to Problem 2 is obtained. In Section 5, we present numerical results demonstrating the effectiveness of the convex initialization to the gradient search.

\subsection{Convex Formulation} \label{Section_Conv} 
Problem \ref{Simp_Problem} is not  convex.
 We show that, by making further simplifying approximations, the problem can be recast into a convex optimization problem. %
In the first step we propose the following parametric CDF approximations. For $m=1,...,M$,
   \begin{align} \label{CXmodel}
{F}_{m}(\delta)\approx\widetilde{F}_{m}(\delta)={c_m\delta^{p_m}}+1-c_m,\quad 0<c_m\leq1, \quad p_m>0,\; 0\leq\delta\leq1,
 \end{align}
where  $p_m$ and $c_m$ are  model parameters obtained by regression. In Section~\ref{Section_USim}, we investigate the ability of the above approximations to represent client RC distributions.

Next, we further simplify the  outage probability constraints. We use the following simpler model \cite{Bakhshali_QBSC12} for the outage probability in order to estimate $N_l$ for each layer:
\begin{align}
    N_l &\approx\frac{S_l+\log_b {P^l_{\text{out}}/a}}{\delta_{l}}, \qquad 0<P^l_{\text{out}}\leq a, \label{N_vs_delta}
\end{align}
where $a$ and $b$ are obtained from the decoding failure probability function of the rateless code \eqref{Erasure}.
Using this the bandwidth constraint becomes 
\begin{align}
   \sum_{l=1}^L \frac{S_l+\log_b {P^l_{\text{out}}/a}}{\delta_{l}}\leq N_{\text{max}} , \qquad 0<P^l_{\text{out}}\leq a. \label{N_vs_delta}
\end{align}
After introducing a parameter transformation $\theta_{l}={1}/{\delta_{l}}, \forall l$, we obtain
\begin{problem}{(Convex formulation)}
\label{final_Problem_conv}
\begin{align}\nonumber
 \ \vspace{-.4cm} & \displaystyle  \min_{\{\theta_{l}\}_{l=1}^L} \sum_{m=1}^M \sum_{l=1}^{h_m} \hat{\alpha}_{m,l} \widetilde{F}_{m}({1}/{\theta_{l}}) \vspace{.2cm} \nonumber
\end{align}
\hspace{1cm}\emph{subject to}
\begin{align}\nonumber
                             \text{UEP Constraints:\quad} &		\theta_{l+1}-\theta_{l}\leq0, \qquad l=1,...,L-1,\\\nonumber
														&  \theta_{L}\geq1, \\													\nonumber
																\text{BW Constraint:\quad} & \displaystyle \sum_{l=1}^{L}{(S_l+\log_b {P^l_{\text{out}}/a})}\theta_{l} \leq N_{\text{max}}.
\end{align}
\end{problem}
\noindent We prove that Problem 3 is convex  in the Appendix. In Section~V we examine the three problem formulations numerically in different application scenarios and assess their accuracies.

\section{Utility smoothing} \label{Section_CliDis}
Source rate and/or service bandwidth fluctuations across consecutive video segments could result in  variations of the optimized MNRCs. Hence, clients with RCs close to the MNRCs may experience quality variations across successive segments. One may encode video segments of longer durations to reduce rate fluctuations at the cost of additional server/client-terminal complexity, memory requirements, and delay \cite{Buffer1, Buffer2}. Below, we reformulate our problem to include suppression of client dissatisfaction due to quality variations.

Major  quality variations are due to unwanted switchings between different layers. This mainly results from the client's RC crossing the MNRC of a layer subscribed by the client.
For example, if a client's RC is always above the MNRC for the base layer, no frame dropping would occur (within the statistical assurance of the base layer outage probability constraint). Below, we extend our problem formulation to  include suppression of MNRC variation.   Numerical results shown later demonstrate the effectiveness of the suppression in reducing  quality switchings, and more specifically, base-layer outage occurrences.

 Let us assume that the client RC distributions do not change significantly across consecutive video segments, i.e., $F_{m}^{(k)}(.)\approx F_{m}^{(k-1)}(.), \forall m$, where $k$ is the video segment index. %
Similar to \eqref{alpha_u}, we define the incremental dissatisfaction coefficients $\beta_{m,l}\geq0$ to  model the client disappointment for not decoding layer $l$ of the current video segment that was successfully decoded previously. Consequently,  the disappointment of a class $m$ client who enjoyed layer $l$ of the previous video segment but can only decode the current video segment up to a lower layer $\hat{l}<l$ is proportional to $\sum_{j=\hat{l}+1}^{l}\beta_{m,j}$.
Using $\beta_{m,l}$, and considering the non-decreasing property \eqref{UEP_CH5} of the MNRCs $\delta_l$, the combined  client dissatisfaction due to MNRC fluctuations is expressed by
\begin{align}  \label{Dis_term}
\mathcal{D}^{(k)}=\sum_{m=1}^M\sum_{l=1}^{h_m}\hat{\beta}_{m,l}[F_{m}(\delta_{l}^{(k)})-F_{m}(\delta_{l}^{(k-1)})]\mathcal{I}(\delta_{l}^{(k)}\geq\delta_{l}^{(k-1)}),
\end{align}
where,  $\hat{\beta}_{m,l}=\pi_m {\beta}_{m,l}$, and  $\delta_{l}^{(k-1)}$ and $\delta_{l}^{(k)}, l=1,...,h_m, \forall m$ are the MNRCs for the previous and the current video segments, respectively. Subtracting  $\mathcal{D}^{(k)}$ from the total utility in \eqref{equ:ServTot} to instrument a variation-induced penalty term leads to the  following optimization problem.
\begin{problem}{(Dynamic optimization)} \label{Problem_NEW}
\begin{align} \nonumber
\ & \hspace{-2cm}\displaystyle \min_{\{\delta_{l}^{(k)}\}_{l=1}^L}  \; \ (1-\lambda)\mathcal{D}^{(k)} +\lambda\sum_{m=1}^M \sum_{l=1}^{h_m} \hat{\alpha}_{m,l} F_{m}(\delta_{l}^{(k)}) &\\ \nonumber
\textrm{\emph{subject to}}  \\[.4cm] \nonumber
	                      \ \text{UEP Constraints:\quad} &  \delta_{1}^{(k)}\ge0, \\ \nonumber
														&		\delta_{l}^{(k)}-\delta_{l+1}^{(k)}\leq0, \qquad l=1,...,L-1,\\\nonumber
														&  \delta_{L}^{(k)}\leq1, \\													\nonumber
                        %
												\ \text{BW Constraint:\quad} & \displaystyle \sum_{l=1}^{L} N_l^{(k)} \leq N_{\text{max}}^{(k)}.
\end{align}
\end{problem} \noindent
$0\leq\lambda\leq1$ effects a balance between the two utility loss terms. A small $\lambda$
 tends to prevent the MNRCs from increasing excessively across two consecutive  video segments. However, longer term gradual increase of the MNRCs due to variations of the RC distributions $F_{m}(.)$ and source bit rate is still possible. However, an exceedingly small $\lambda$ may significantly reduce the overall utility provided to the clients. Hence, a judicious choice of $\lambda$ would avoid letting clients with the worst channels from unduly influencing the solution.

\section{Utility optimization using a perceptual quality metric}
\label{Section_Util}

The proposed multicast optimization scheme can be tailored to fit different application scenarios. Here, we aim to maximize the clients' subjective viewing  experience by setting the marginal utility parameters $\alpha_{m,l}$ using a perceptual quality model that was developed using subjective-viewing test results \cite{NMOS_Xue1, NMOS_OU2}.
Although peak signal-to-noise ratio (PSNR) \cite{VideoQ_Survey} has been widely used as a measure of video quality, low correlation between PSNR and video quality ratings provided by human viewers---commonly reported as  mean opinion scores (MOSs)---is reported. The shortcomings of PSNR  are  more pronounced when comparing  video playback at different spatial and temporal resolutions.  %
More versatile objective quality measures have been proposed as estimates of subjective quality ratings. 
The objective video quality metric introduced in  \cite{NMOS_Xue1} and \cite{NMOS_OU2} provides a normalized MOS (NMOS) that can be used to quantify the quality between different spatial, temporal and quantization resolutions
\begin{align} \nonumber
\mathrm{NMOS} (s,f, \text{PSNR})&= \left(\frac{1-e^{-b_s\frac{s}{s_{\text{max}}}}}{1-e^{-b_s}}\right)\left( \frac{1-e^{-b_f\frac{f}{f_{\text{max}}}}}{1-e^{-b_f}} \right)\\
 & \qquad \times \left(1-\frac{1}{1+e^{0.34(\text{PSNR}-b_p)}} \right). \label{NMOS}
\end{align}
Here $s$ and $f$ represent the number of pixels and frame rate respectively, while $s_{\text{max}}$ and $f_{\text{max}}$ are  their maximum values.  $b_s$, $b_f$ and $b_p$ are  model parameters that depend on the video content  \cite{NMOS_Xue1, NMOS_OU2}. This NMOS model is conveniently used to illustrate the method proposed herein. More elaborate quality estimation methods such as the video quality metric (VQM) \cite{VQM} algorithm may be advantageously employed. We should mention that a slightly advanced version of the NMOS model used in this work was published in \cite{NMOS_TIP_2014}.

As an illustration, consider a scenario wherein the highest video layer successfully recovered by a terminal has a spatial resolution lower than the playback capability; specifically, a HD terminal receiving a SD video. The terminal may display the SD video as received in the middle of the HD display or adapt the video to the display by upsampling. The  perceptual quality metric in \eqref{NMOS} is used as a yardstick to compare the perceptual effects of various possible adaptations. 
$\mathrm{NMOS}_{m,l}$, non-decreasing with layer index $l$, represents the highest NMOS corresponding to the best possible adaptation---within the capabilities of the class $m$ terminals---that can be performed on the media up to layer $l\leq h_m$. Recall that $h_m$ is the highest layer of the video stream that class $m$ terminals can potentially decode.  %
Furthermore, we may also model client playback preferences  that can be set independently of the achieved video quality. For example, a certain application may require the spatial resolution not to be lower than some specific level. We may use preference weights $0\leq W_{m,l}\leq1$, non-decreasing with respect to index $l$, to  map NMOS to multicast utility while accounting for clients' playback preferences. If class $m$ users are unwilling to settle for media playback at any layer $l<h_m$, then $W_{m,l}=0  \ \; \forall l<h_m$.
Thus, we define our utility-rate function as
\begin{align}
\mathcal{U}_m(R_l)=W_{m,l}\mathrm{NMOS}_{m,l}, \label{service_mod2}
\end{align}
which can be applied to \eqref{alpha_u} to calculate the marginal utility coefficients $\alpha_{m,l}\geq0$.

\section{Numerical Simulations}
\label{Section_USim}

In the following, we evaluate the performance of  the proposed optimization scheme when applied to multicasting a video sequence with $L=3$ layers to heterogeneous clients. Bit stream parameters for three H.264/SVC encoded video sequences are summarized in Table \ref{H264RD}. $S_l$ denotes the number of source symbols in each layer over a $T_{\text{seg}}=1$ second time segment and each symbol comprises 50 bytes.

The bit rates given are for each layer. 
The OP constraints $\mathbf{P}_{\text{out}}=\{P_{\text{out}}^1, P_{\text{out}}^2, P_{\text{out}}^3\}=\{10^{-4},  4\times10^{-4},5\times 10^{-4}\} $ are enforced.
An equal error protection (EEP) scheme  is used as the base-line for the performance comparison. In the EEP scheme, the transmission resources allocated to each media layer is proportional to the relative size of that layer in the source bitstream, i.e.,
\begin{align} \nonumber
N^{{e}}_{l}=\frac{{N_{\text{max}}}S_l}{\sum_{k=1}^L S_k} \qquad l=1,...,L.
\end{align}
We use the following metrics  to evaluate the performance gain and efficiency of different schemes, respectively,
\begin{align}
\textrm{$\eta_{{}}\uparrow$}\triangleq\frac{{U}-{U_{{e}}}}{{U_{{e}}}} \textrm{\%}, \qquad \textrm{$\varepsilon$}\triangleq\frac{{U}}{{U_{{opt}}}} \textrm{\%},
\end{align}
where ${U}$ is the utility delivered to the clients, ${U_{opt}}$ is the maximum attained by using the optimal MNRCs, and ${U_e}$ corresponds to the utility of the EEP scheme.  
 The interval $0<\delta\leq1$ is partitioned into small sub-intervals and an exhaustive search is performed to find the MNRCs $\delta_{l}, l=1,\dots,L$ and subsequently ${U_{opt}}$. Nevertheless, this process could be computationally expensive for large number of sub-intervals and source layers. To obtain a sub-optimal solution with much lower complexity, first, the convex problem (Problem 3) is solved. Next, a constrained gradient descent (GD)  algorithm is deployed to solve the simplified formulation formulation (Problem 2), using the convex solution as a starting point. The performance measures of these two  solutions are superscripted ``CV'' and ``GD'' respectively.
The multicast clients may experience a wide variety of channel conditions depending on fading and their distance to the transmitting station \cite{RC_Wireless_Wang}. For wide-area cells, the range of channel qualities can be expected to be broader than reported in \cite{RC_Wireless_Wang}.  Thus, the uniform distribution and truncated Gaussian mixtures in Fig. \ref{fig:DIST_M1} are selected to reflect distinct types of client RC statistics with different balances between the number of clients with poor and good channels. 1,000 clients are considered for these scenarios.  In the multi-class scenario, each class inherits a portion of clients based on the priors $\pi_m, m = 1,...,M$. Next, samples of the client reception coefficients (RCs) are generated
for each distribution.  

\subsection{Single-class Scenario}
In this scenario, all clients  are assumed to be capable of decoding all three layers. Hence, $M=1$ and $h\triangleq h_1=L=3$.
 Table \ref{Table_Results_Single} exhibits the optimization results for $N_{\text{max}}=13,000$ symbols.

The performance metrics are evaluated over four crafted utility settings. 
On average, the  proposed optimization  manages to increase the utility by a factor of more than 2  compared to the EEP solution. The EEP solution is highly inefficient when the majority of clients experience poor channels, as in the $\Delta$-III distribution in Fig. \ref{fig:DIST_M1}. Note that the solution of the convex optimization yields an average efficiency of $95.25 \%$. Adding the GD search increases the efficiency to $99.50 \%$. 
 The optimization results as well as  the solution of the EEP approach for different service bandwidth constraints $N_{\text{max}}$ are depicted in Fig. \ref{fig:SingClass_N}. The metric values are averaged over the four distributions and utility settings (the 16 cases in Table \ref{Table_Results_Single}).

Due to the high efficiency of the initial convex solution, the GD step could be omitted in order to reduce computation without significant performance penalty. 

\subsection{Multi-class Scenario}

Next, we consider a scenario with $M=2$ client classes. The class 1 clients with CIF resolution displays may only decode the base layer and the first enhancement layer, i.e., $h_1=2$. The clients in class 2 have 4CIF resolution displays and decoders capable of decoding the entire video stream, i.e., $h_2=3$. %
The four sample distributions in  Fig.~\ref{fig:DIST_M1} are used to model the client RC distributions of both client classes, resulting in 16 possible distribution pairings. For each pair of distributions, the simulation is performed with different prior values. 
 The utility parameters are obtained using the perceptual quality metric in \eqref{service_mod2}. The preference parameters are assumed to be $W_{m,l}=0.9^{h_m-l}, l\leq h_m$. 
 The NMOS parameters for  the test video sequence are extracted from \cite{NMOS_Xue1} and \cite{NMOS_OU2}. 
The simulation results are shown in Table \ref{Table_Results} in terms of metric values averaged over the 16 pairings. 

On average, the initial allocation provided by the convex approximation achieves $97.57 \%$ efficiency. Using the GD algorithm, the efficiency is increased to $99.80 \%$. Similar to the single-class scenario, most of the potential performance gain can be obtained using the  convex optimization. %

\subsection{Reduced-feedback Scenario}
It is worthy to investigate the optimization performance when client RC statistics are collected only from  a  portion of the multicast clients. Limiting channel state information feedback could be an effective measure against feedback implosion at the server and for maintaining a low error rate for a multiple access feedback channel. %
For all 16  pairings of the  RC distributions in Fig. \ref{fig:DIST_M1} for the class 1 and 2 clients, an ensemble of size $n_m=1,000, m=\{1,2\}$ samples are drawn from each distribution to represent 1,000 clients in each class ($\pi_1=\pi_2=1/2$). The performance is evaluated as a function of the fraction of clients from each class that successfully send their RC and media player capability information to the server---in terms of client-to-server feedback ratio (CSFR), $0\leq \text{CSFR}\leq1$. This experiment is repeated 100 times for every CSFR and distribution pairing to ensure accuracy, especially for small CSFR values.  
The histograms of the received RC feedback messages are used as  estimates of the actual class RC distributions, and employed in the optimization. 
The performance is compared to the scenario in which full knowledge of all client RCs is revealed to the server, i.e., all clients successfully feed back their RCs to the server (CSFR=1). For each CSFR,  the performance metrics of all three tested video sequences are combined (4800 simulation runs per CSFR) and the results are illustrated in Fig.~\ref{RedFeed_fig}.

The proposed optimization demonstrates good tolerance to limited RC feedback. Optimization based on RC feedback from only 5\% of the clients still provides  performance close to 100\% feedback. Both convex optimization and the GD algorithm maintain their performance in the limited feedback regime. For a smaller pool of 100 clients per class, the CFSR needed goes up to about 20\%.  However, the small number of  feedback clients, 20 in this example, should be manageable. We believe this robustness comes from the ability of the parametric CDF in \eqref{CXmodel} to capture the general characteristics of the client RC distributions. 
\subsection{Variable Rate Source Scenario}
Performing a resource allocation optimization repeatedly for each video segment means that computation intensity depends on segment duration  $T_{\text{seg}}$. One way to reduce computation is to use a large $T_{\text{seg}}$ though $T_{\text{seg}}$ may be limited by other considerations such as media bit stream access and formatting requirements. Another way is to  optimize the video less frequently by using longer term statistics. In this section we aim to quantify the performance penalty incurred  when the optimization uses longer-term statistics as compared to segment by segment optimization.
Note  that a video bit stream may exhibit large bit rate variations due to intra-coded frames. Longer video segments can reduce the rate fluctuations at the cost of additional buffering.
Let us consider  $R_l^{(k)}=S_l^{(k)}/{T_{\text{seg}}}$ as the source rate for layer $l$ of  video  segment $k$ with duration $T_{\text{seg}}$ seconds and  $S_l^{(k)}$ source symbols. We model the source bitstream variations across different segments by
\begin{equation} \label{VarS_EQ}
S_l^{(k)}=S_l (1+ {\gamma}_l^{(k)}),
\end{equation}
where $S_l$ is the average length of layer $l$ obtained from Table~\ref{H264RD}, and ${\gamma}_l^{(k)}, l=1,..,L, \forall k$ are $L$ independent and identically distributed uniform variables with  support $[-\gamma_{\text{\text{max}}},\; +\gamma_{\text{\text{max}}}]$. 
For the special case  $\gamma_{\text{max}}=0$, the source becomes a constant-rate source (CRS) and the optimal allocation is independent of any particular video segment $k$ provided that the service bandwidth, the utility coefficients, and client RC distributions are fixed. 
The following efficiency measure quantifies the performance penalty due to performing the resource allocation optimization using average statistics,
\begin{equation}
{\Large\varepsilon_{_{CRS}}}= \langle {U_{CRS}}^{(k)}/{U}^{(k)}\rangle.
\end{equation}
Here, $\langle.\rangle$ denotes  averaging over segments.  ${U_{CRS}}^{(k)}$ is the utility achieved  for the $k^{\text{th}}$ video segment when the resource allocation optimization is performed only once based on average rate-distortion statistics. Conversely, ${U}^{(k)}$ is the maximum attainable utility when a separate resource allocation optimization is conducted for each video segment. 
The maximum number of transmitted packets $N_{\text{\text{max}}}$ and the client RC distributions are assumed to remain unchanged during the entire multicast. 
For every  $\gamma_{\text{max}}$ and 16 pairings of the candidate distributions, 100 samples of ${\gamma}_l^{(k)},\forall l$ are generated to represent variable source rates for 100 video segments.
The results are plotted in Fig.~\ref{Fig_VSR} as a function of the max-to-min rate ratio (MRR) for the video  rates generated by \eqref{VarS_EQ} where $\text{MRR} \triangleq\frac{1+\gamma_{\text{max}}}{1-\gamma_{\text{max}}}$.

As expected, optimization based on long-term statistics results in lower efficiency. However, the performance penalty is moderate since ${\Large\varepsilon_{_{CRS}}}$ remains at above 90\% efficiency even for a rate variation as large as $\text{MRR}=19$. 
We should mention that the efficiency ${\Large\varepsilon_{_{CRS}}}$ of the EEP solution  remains  below 65\% for $N_{\text{max}}=15,000$ and $N_{\text{max}}=19,000$, respectively, reconfirming the poor performance of the EEP solution for quality-aware multicast transmission.
\subsection{Multi-Segment Quality Smoothing}
In this scenario the proposed dynamic utility maximization (Problem 3) which penalizes quality fluctuations for clients with marginal RCs is studied. We consider 9 consecutive segments of an H.264-SVC coded multilayer ($L=3$) \emph{Crew} video sequence, each segment containing 32 frames with QCIF and CIF spatial layers, and the GOP size is 16 frames with one intra coded frame starting each GOP. 
The base-layer embeds the QCIF resolution with frame rate of 15 frames/s. The quantization parameter (QP) for the base layer is set to 44. The first enhancement layer increases the frame rate from 15 to 30 frames/s and additionally provides a better quantization resolution with QP=32. Finally, the last enhancement layer embeds the CIF resolution with frame rate and QP identical to the previous layer. The NMOS model parameter values for this video sequence are $b_f=7.23$,  $b_s=3.49$, and  $b_p=29.68$ dB \cite{NMOS_Xue1,NMOS_OU2}. We observed that the encoded sequence provides nearly steady PSNRs across the video segments. The achieved PSNRs are 30.5 dB, 35.1 dB, and 35.2 dB for the base layer and the enhancement layers, respectively. Based on these PSNR values and the model parameters \cite{NMOS_Xue1,NMOS_OU2}, the average NMOS values are 0.31 for the base layer, 0.48 for the second layer, and 0.86 for the third layer. These scores manifest a peak variation of less than 5\% across different video segments.
 Two client classes  ($M=2$) with equal population size ($\pi_1=\pi_2=0.5$) are assumed. $\Delta$-II and $\Delta$-IV from Fig.~\ref{fig:DIST_M1} model the RC distributions of the class 1 and 2 clients with QCIF and CIF screen resolutions, respectively. The video decoders of  the class~1 clients are assumed to be capable of decoding the base-layer as well as the first enhancement layer while class 2 clients are capable of decoding all video layers.

Here, we aim to optimize the provided utility under the constraint of limited service bandwidth $N_{\text{max}}$. Additionally, 
 failure in decoding the base layer is considered unacceptable for both classes. Therefore, we set the dissatisfaction coefficients $\beta_{m,1}=1 \; \forall m$ and the rest of the dissatisfaction coefficients to zero. The server is assumed to transmit $N_{\text{max}}=11,000$ symbols for each segment, where each symbol consists of 16 bytes.

The video segment size, 
the optimized utility for various values of $\lambda$, and the optimized MNRCs $\delta_l^{(k)}, l=1,...,3$ are plotted as a function of the segment index $k$ in Fig.~\ref{Fig_Dynamic}. This video sequence exhibits a significant rate increase at the 4th segment. This raises the MNRC for the base layer  $\delta_1^{(k)}$ when the quality fluctuation suppression term is nulled ($\lambda=1$). 
By increasing $\lambda$,  the optimization increasingly penalizes solutions that allow the base layer MNRC to increase. Hence, the portion of clients that face temporal outage is reduced and a more stable visual experience is provided. This is reflected in lower $\delta_1^{(k)}$ values with smaller variations. Given a fixed service bandwidth and considering the fact that $\beta_{m,l}=0$ for $l\geq2$, the reduction  in the  MNRC fluctuations for the base layer comes at the cost of increased variations of the  MNRCs for the enhancement layers, as reflected in the $\delta_2^{(k)}$ and $\delta_3^{(k)}$ traces in Fig.~\ref{Fig_Dynamic}(d-e). 
 Note that the achieved utility is closer to the upper-bound  ${U_{\text{max}}}$ for the video segments with fewer source symbols. ${U_{\text{max}}}$ depends on the video content and its viewing quality but not the source rate. However, the gap between the achieved utility and ${U_{\text{max}}}$ depends on the portion of clients who are unable to receive the video layers they desire.  Hence, for a constant $N_{\text{max}}$, increasing the source rate widens the gap, in proportion to the distribution of clients with marginal RCs. Additionally, we observe that the penalty term with different weights $(1-\lambda)$ hardly affects the utility traces except for the 4th segment that contains the sudden rate increase.%

For the particular choice of $\beta_{m,l}$  values in this scenario, the  average fraction of clients that successfully enjoys the base-layer in one video segment but fails to decode the base-layer in the next segment can be obtained by the averaging the dissatisfaction measure  \eqref{Dis_term} over the video segments $\overline{\mathcal{D}}=\langle\mathcal{D}^{(k)}\rangle$. This metric is the marginal probability that a satisfied user encounters frame drops or freezes in the next video segment. Table \ref{Dynamic} illustrates $\overline{\mathcal{D}}$ for various service bandwidths and $\lambda$. As expected, higher bandwidth and smaller $\lambda$ both contribute towards a more stable video quality experience.
Table \ref{Dynamic} also provides data for $\mathcal{Z}$ which measures the percentage of clients that experience outage in decoding the base-layer at least once during the 9 video segments. Due to the client RC  distributions modeling a significant portion of clients with poor channels, and a notable rate increase beyond  the 4th segment, there is always a portion of clients that experience outage for a constant $N_{\text{max}}$. When the variation suppression term $\mathcal{D}$ is disabled, the outage percentage remains stubbornly high even as the service bandwidth is substantially increased. However, a lower outage rate  $\mathcal{Z}$ is attainable by increasing the penalty weight $1-\lambda$. If segments 4 to 9 are excluded from  the statistics for $N_{\text{max}}=11,000$, $\mathcal{Z}$ is reduced from 16.03\% to 3.24\% for $\lambda=1$. However, for $\lambda=0.3$ exclusion of those segments  reduces $\mathcal{Z}$ slightly from 3.6\% to 2.85\%. This signifies the performance of the proposed dynamic optimization in reducing the sensitivity of client dropout to high rate video segments.

The outage statistics  based on the $\mathcal{Z}$ measure for the EEP solution is 19\% to 30\% higher than the proposed  dynamic optimization. Fig.~\ref{burst_Dynamic} provides the outage burst length statistics assuming that client  channel quality is unchanged during the transmission of the 9 video segments. The results are normalized to the number of  maximum length outage incidents for the EEP scenario. It is clear that the proposed optimization significantly reduces the number of outage incidents.%

Furthermore, we investigate the performance of the proposed algorithm for a client with time-varying RC. We consider a client with a poor average RC $\overline{\delta_c}=0.2$. Based on  the MNRC $\delta_1^{(k)}$ traces in Fig.~\ref{Fig_Dynamic}(c), the viewing experience of this client would be disturbed by base layer outage. We use truncated normal distributions with mean $\mu=0.2$ and different standard variations $\sigma$ to model the probability distribution of its RC during the transmission of all 9 segments. Examples of these distributions are depicted in Fig~\ref{RC_Dynamic}. We calculate the frame freeze rate (FFR), defined as the percentage of  frames not received and may be  replaced by the last decoded frame.  The results are depicted in Fig.~\ref{FFR_Dynamic}. Using the proposed optimization, the FFR is reduced by as much as 11\% and 7\%  for  narrow RC distributions ($\sigma<0.02)$ and wide distributions ($\sigma>0.02$), respectively. Note that the FFR for the  EEP solution is more than 99\% for this client due to significantly higher values of the corresponding $\delta_1^{(k)}$ traces.

In practice, it may be possible to vary the service bandwidth $N_{\text{max}}$ with the source symbol rate.  For instance, a server simultaneously serving multiple independent video streams can exploit a well-known advantage offered by statistical multiplexing: the total source rate fluctuates far less than the individual source rates. In such case, allowing $N_{\text{max}}$ to vary, in conjunction with the proposed method, would enable suppression of outage to negligible levels. The MNRCs can also be transmitted as a side information with negligible cost. Therefore, a client can select a video layer for playback whose MNRC is at a safe margin below the client's RC. The client may use the MNRCs for the previous segments as input to an algorithm that selects the actual enhancement layers for decoding and display, with the aim to produce the best viewing experience. MNRC smoothing helps the algorithm to achieve a good viewing experience.

\section{Conclusions}
\label{Section_Conc}

Considering heterogeneity of client channels and their terminal capabilities, we introduced a  QoE optimization framework for  video multicast that benefits from the flexibility offered by scalable video coding and fountain coding. The client's ability to decode different video quality layers is exploited  to maximize the overall utility of the multicast transmission. Utility is formulated based on a perceptual quality metric that can differentiate between various possible adaptations of a multilayer video stream with a combination of spatial, temporal and granular scalability. The optimization effects a balance between QoS-guaranteed service and best-effort service. Catering to the probabilistic decoding nature of rateless codes, outage probability constraints are applied to guarantee that the video quality layers are received with high level of assurance. Clients that cannot be served meeting  such guarantees may be served with a lower playback quality from the lower video layers. Clients demanding high-quality playback but present in small numbers may be similarly treated. Clients with exceedingly poor channels may be  dropped from the multicast. On the other hand, given a sufficient transmission rate, clients are served the highest quality playback level they desire. 
 The optimization complexity is independent of the number of clients and scales only with the number of client classes. Additionally, a convex optimization approximation is proposed which has shown to attain close-to-optimal performance with even lower computational complexity. The proposed optimization framework is also shown to provide robust performance when limited client feedback information is available. Finally, by introducing a penalty term to the multicast utility, the QoE optimization is extended to suppress  client playback quality variations due to source bit rate and/or service bandwidth fluctuations. Despite the above promising results, a possible future work would be to assess the efficacy of the proposed scheme in more full-fledged application scenarios similar to \cite{wu2013joint}.

\appendix 


\section*{Appendix: Convexity Analysis of Problem 3} 
\label{APP_Conv}
For the convexity analysis we form the Hessian matrix $\mathbf{H}$ from the second derivatives of the cost function with respect to the optimization variables $\theta_{l}, l=1,...,L$, 
\begin{align} \nonumber
\mathbf{H}=\bigg[H_{jk}\bigg]&=\bigg[\frac{\partial^2}{\partial \theta_{j} \partial \theta_{k}} \sum_{m=1}^M \sum_{l=1}^{h_m} \hat{\alpha}_{m,l} \widetilde{F}_{m}({1}/{\theta_{l}})\bigg] \qquad  j,k=1,...,L.
\end{align}
The diagonal elements can be obtained from differentiating \eqref{CXmodel} 
\begin{align} \nonumber
H_{jj}&=\sum_{m=1}^M \hat{\alpha}_{m,j} \frac{\partial^2}{\partial \theta_{j}^2} ({c_m\theta_{j}^{-p_m}+1-c_m})\\
&=\sum_{m=1}^M \hat{\alpha}_{m,j}c_mp_m(p_m+1){\theta_{j}^{-(p_m+2)}}.
\end{align}
Since $c_m,p_m,\hat{\alpha}_{m,j}\geq0$ and $\theta_{j}\geq 1, \  \forall j, m$, we conclude that $H_{jj}\geq0, \forall j$.

Similarly it can be shown that the off-diagonal terms of the Hessian matrix $H_{jk}, j\not=k$ are zero. As a result, $\mathbf{H}$ is positive semidefinite and the cost function is convex \cite{Boyd_Book}.

The UEP constraints in Problem 3 are linear since they are of the form $\theta_{j+1}-\theta_{j}\leq0$ with $j=1,...,L$ and $\theta_{L+1}\triangleq 1$. Furthermore, the bandwidth constraint is also linear. Hence, the constraints form a polyhedron which is a convex set. Since the convexity of the cost function was previously established, Problem 3 is a convex optimization problem.


\begin{backmatter}

\section*{Competing interests}
  The authors declare that they have no competing interests.
%

\section*{Acknowledgements}
This work was supported by the Natural Sciences and Engineering Research Council of Canada.

\bibliographystyle{bmc-mathphys} 
\bibliography{EURASIP_bibtex}      




\section*{Figures}
\begin{figure}[h!]
\includegraphics[width=12cm]{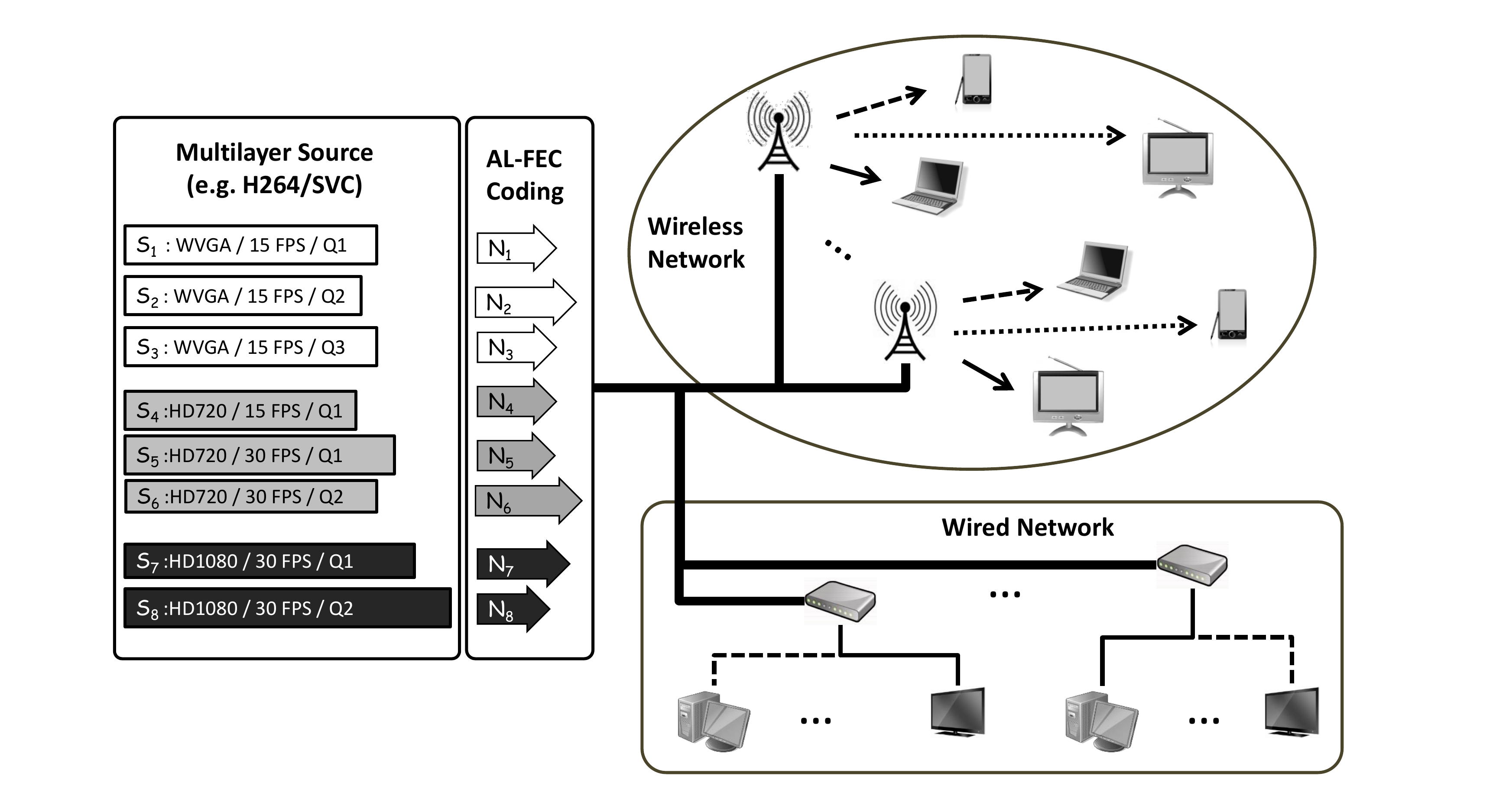}
\caption{\csentence{System setup.} System setup for the proposed rateless-coded based video multicast. }  \label{fig:Scheme}
\end{figure}

\begin{figure}[h!] 
\includegraphics[width=12cm]{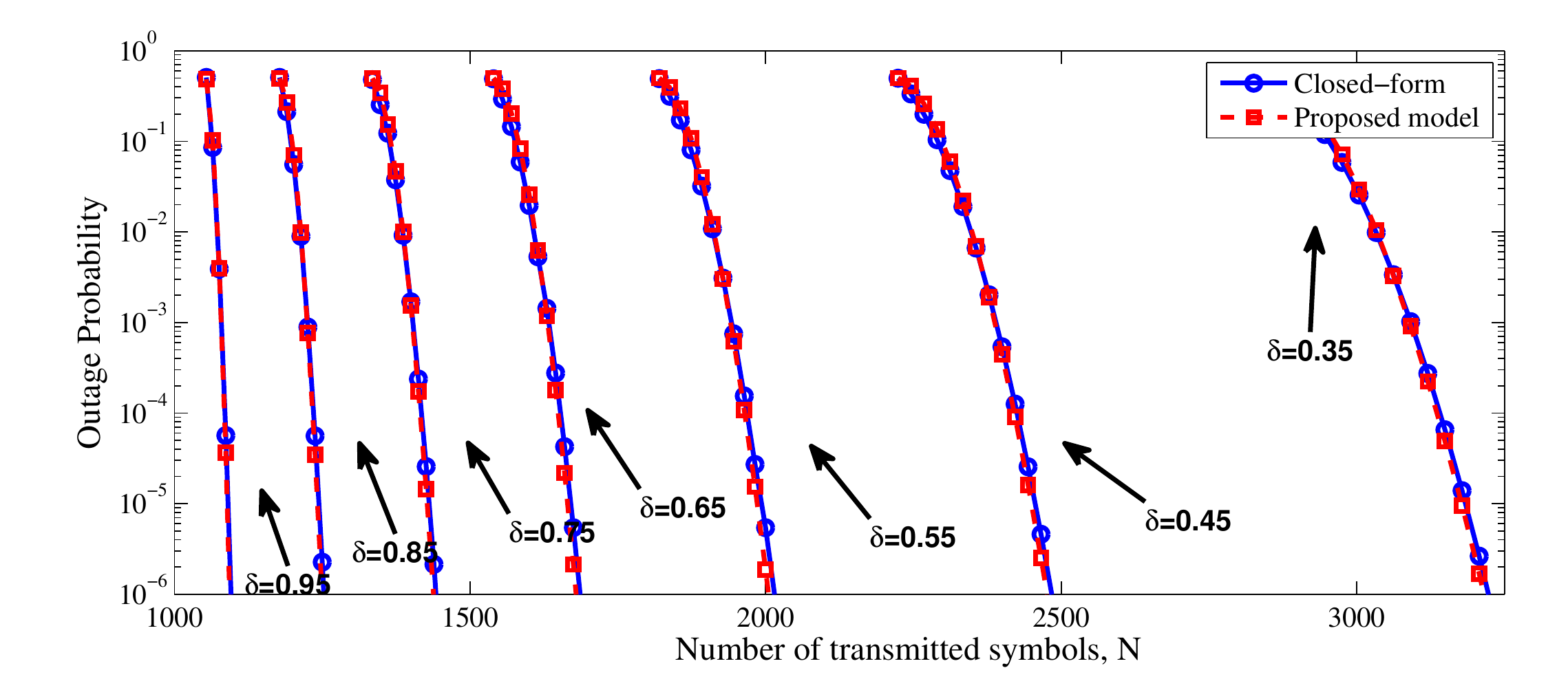}
\caption{\csentence{Comparison between the closed-form outage probability and approximated outage probability model.} Closed-form outage  probability \eqref{Pout1} and the outage probability obtained from the approximated model \eqref{OutMod} as a function of transmitted fountain symbols for a source with size $S=1000$.} \label{OutFig}
\end{figure}

\begin{figure}[h!]%
\includegraphics[width=12cm]{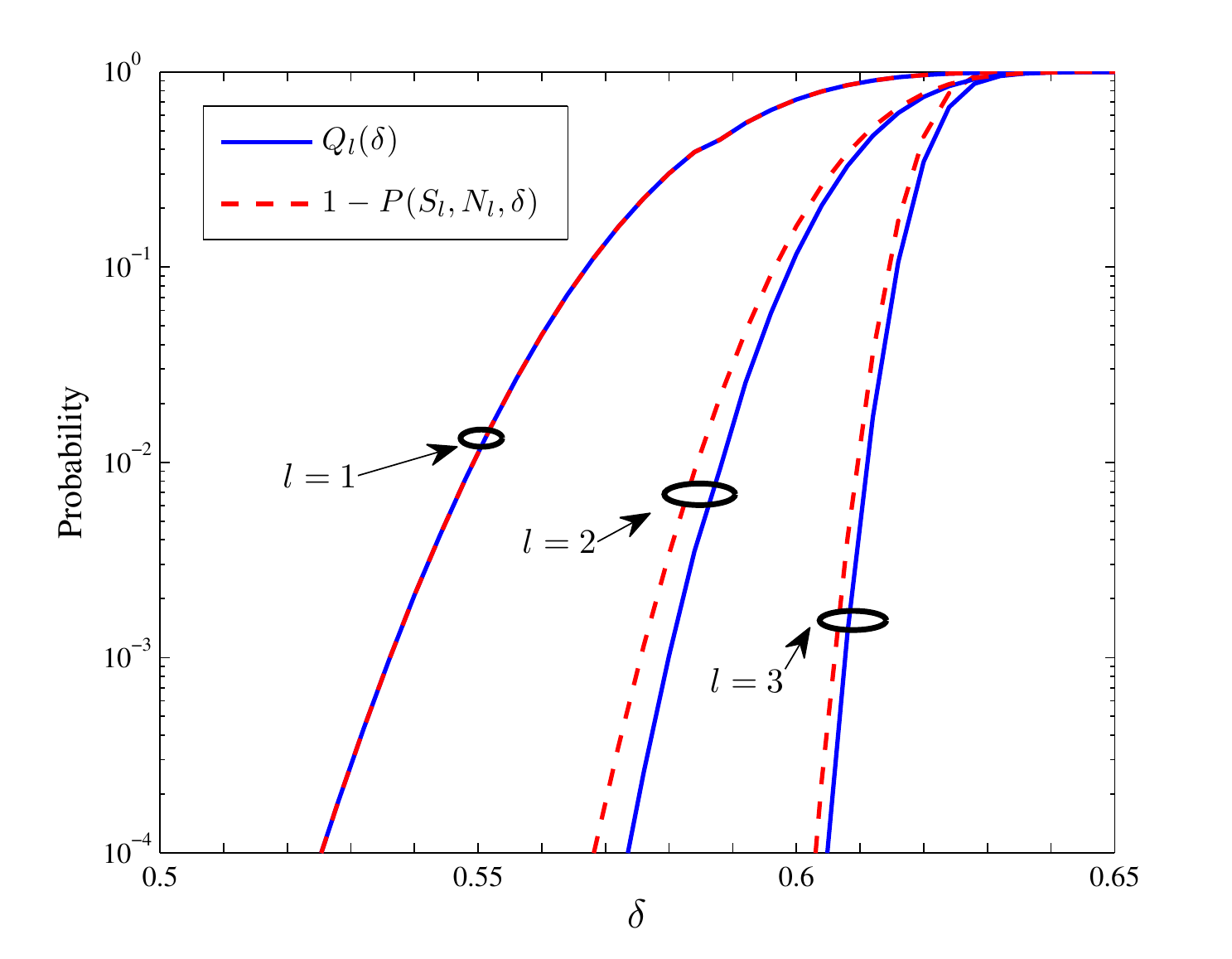}
\caption{\csentence{Outage probability approximation.} A comparison between $Q_l(\delta)$ (solid blue) and $[1-P(S_l,N_l,\delta)]$ (dash red) for the different layers  of a three-layer video stream. $\delta_1=0.526, \delta_1=0.572, \delta_1=0.607$ for outage probability constraints similar to those expressed in Section~5.}%
\label{SimpOutage_fig}%
\end{figure}

\begin{figure}[h!]
\includegraphics[width=12cm]{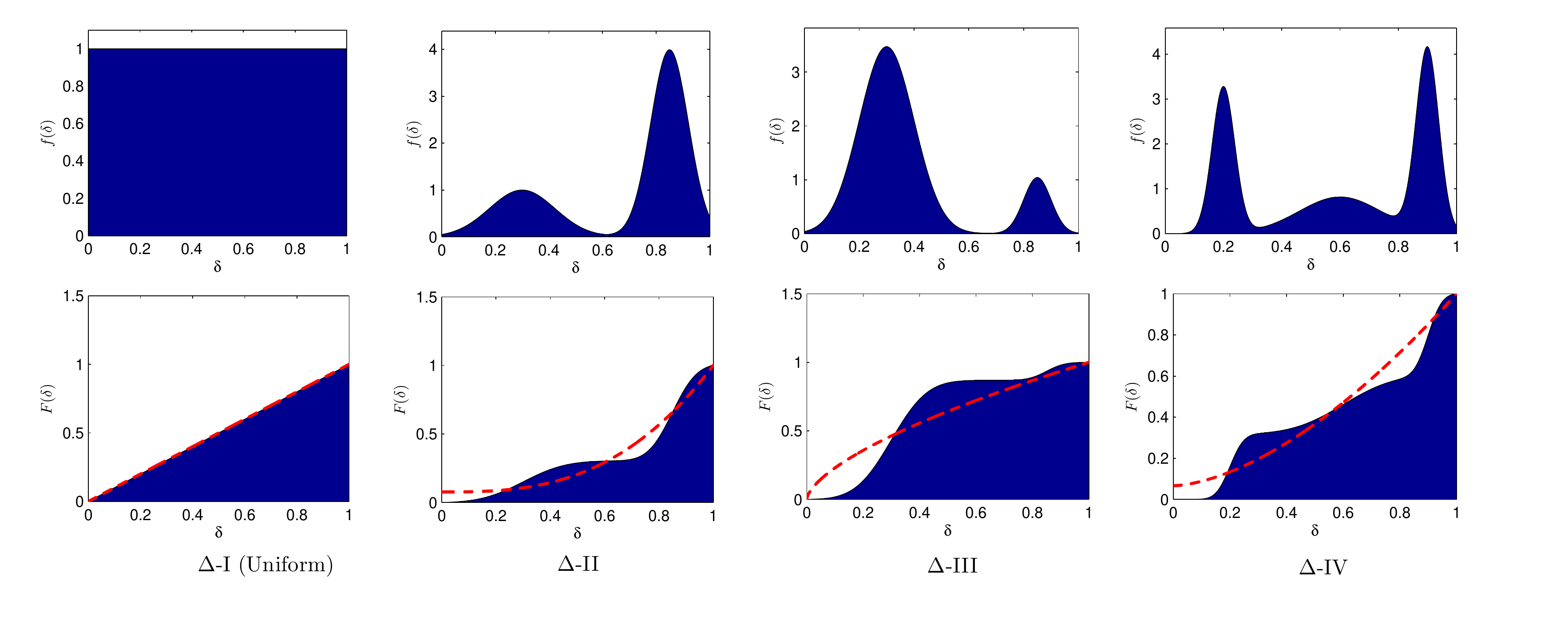}
\caption{\csentence{Class RC distributions}. Prototypical (top) PDFs and their corresponding (bottom) CDFs. Approximated CDFs based on \eqref{CXmodel} are depicted in dash-red.}  \label{fig:DIST_M1}
\end{figure}

\begin{figure}[h!]
\includegraphics[width=12cm]{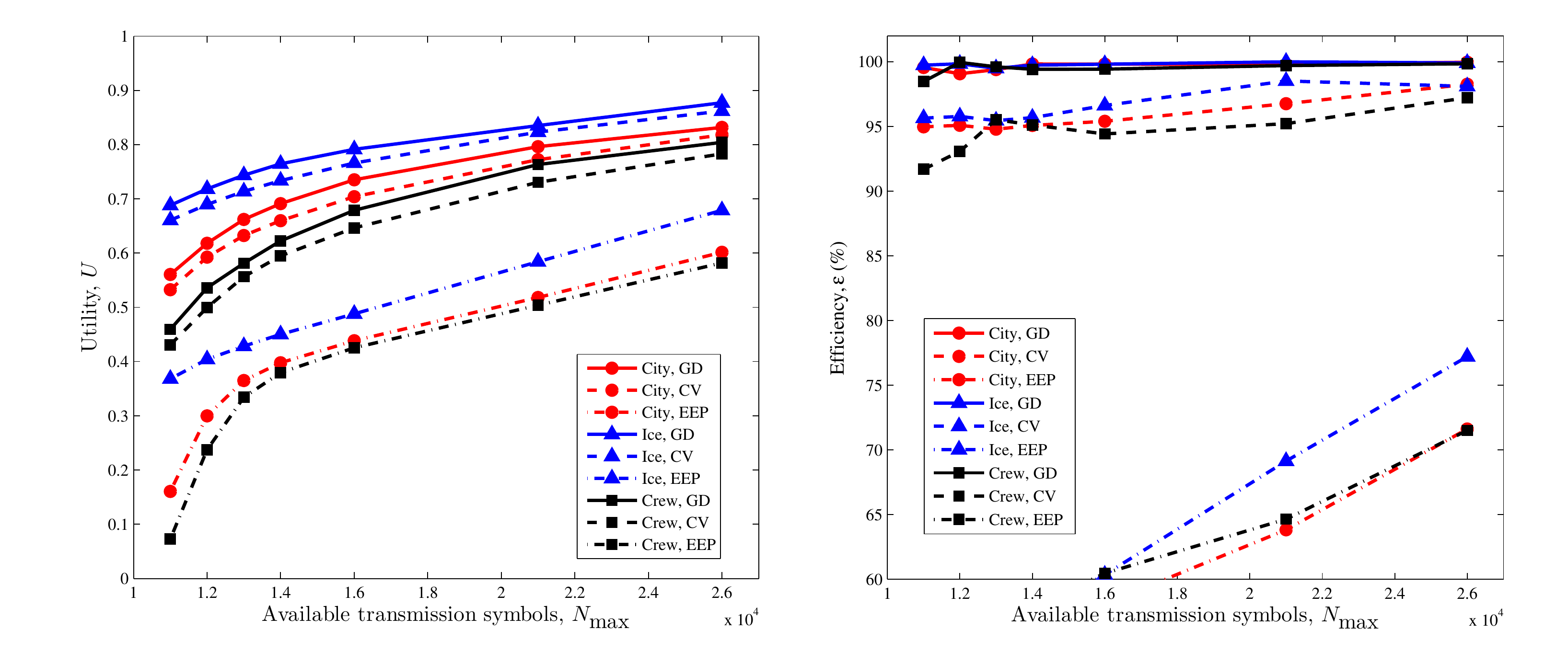}
\caption{\csentence{Single-class optimization results.} (left) Average utility traces and (right)  efficiency of the single-class video multicast optimization for various transmission budgets. }  \label{fig:SingClass_N}
\end{figure}

\begin{figure}[h!]%
\includegraphics[width=12cm]{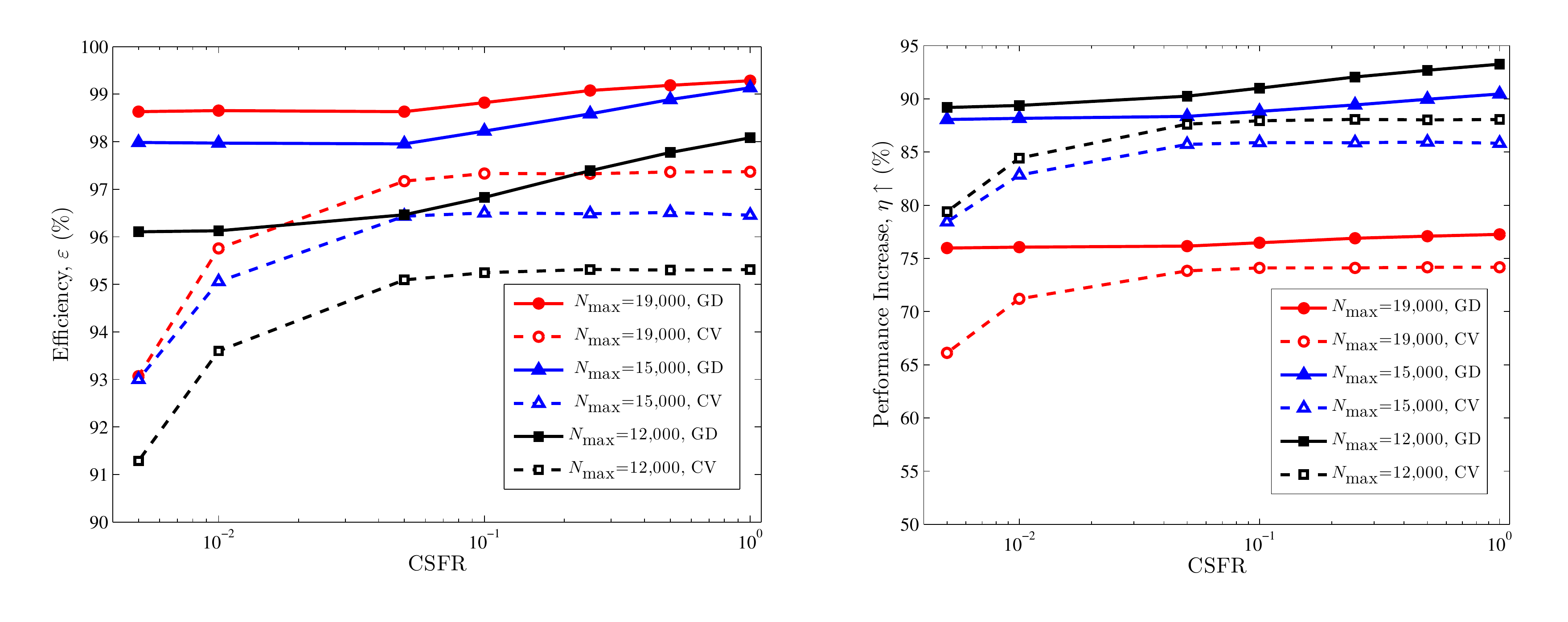}
\caption{\csentence{Optimization results for the reduced-feedback scenario.} Performance of the proposed optimization as a function of the fraction of clients that successfully feed back their channel state information to the server for (dash-line) convex optimization, and (solid-line) GD method.}%
\label{RedFeed_fig}%
\end{figure}

\begin{figure}[h!]%
\includegraphics[width=12cm]{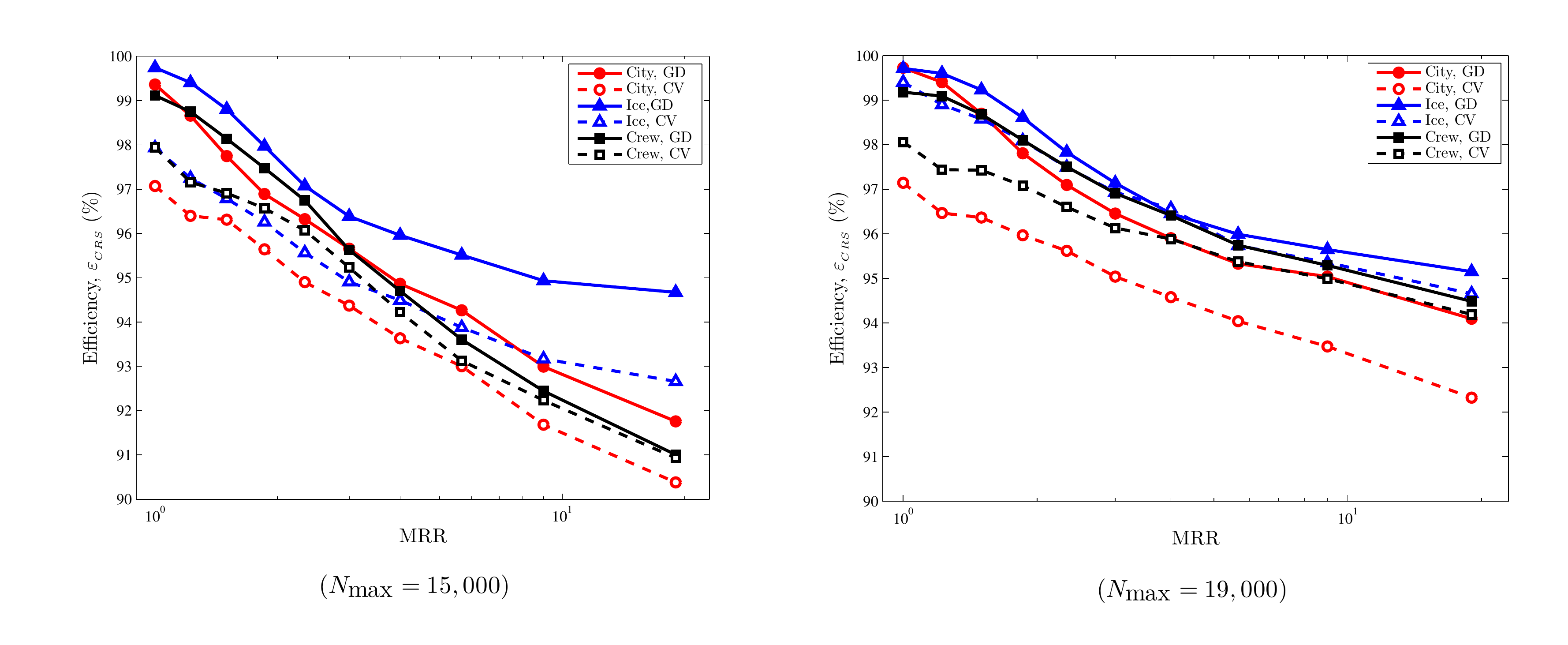}
\caption{\csentence{Long-term statistics based optimization results for a variable-rate source.} Optimization efficiency based on the long-term statistics of a variable-rate source for (left) $N_{\text{max}}=15,000$ and (right) $N_{\text{max}}=19,000$. Sources with different max-to-min rate ratios (MRRs) are emulated by modifying the distribution of ${\gamma}_l^{(k)}$ in \eqref{VarS_EQ}.}%
\label{Fig_VSR}
\end{figure}

\begin{figure}[h!]%
\includegraphics[width=12cm]{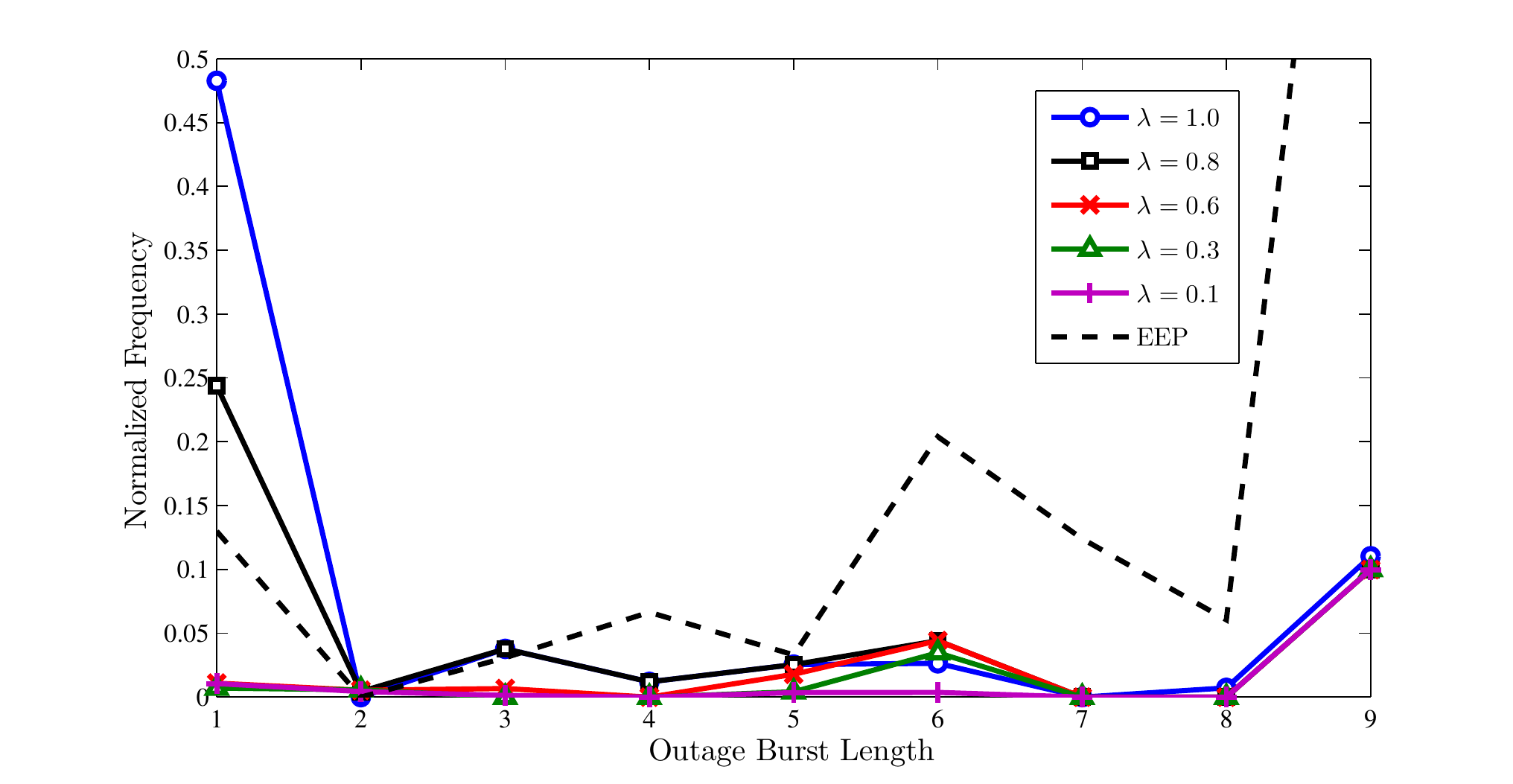}%
\caption{\csentence{Outage burst statistics.} Outage burst statistics for different solutions of the dynamic optimization, $N_{\text{max}}=11,000$ }\label{burst_Dynamic}%
\end{figure}

\begin{figure}[h!]%
\includegraphics[width=12cm]{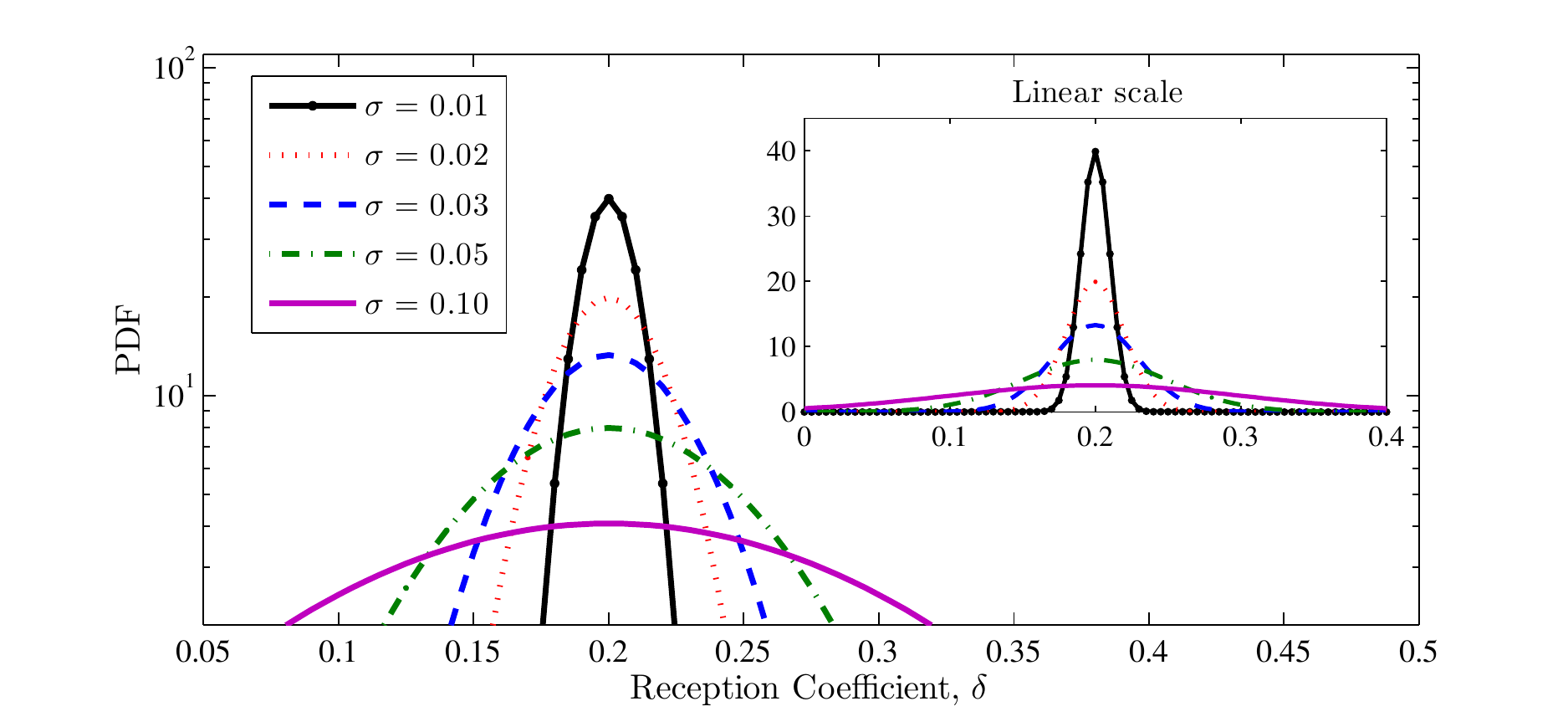}%
\caption{\csentence{RC distribution of a client with dynamic channel.} RC distribution of a client with an average RC $\overline{\delta_c}=0.2$} \label{RC_Dynamic}%
\end{figure}

\begin{figure}[h!]%
\includegraphics[width=12cm]{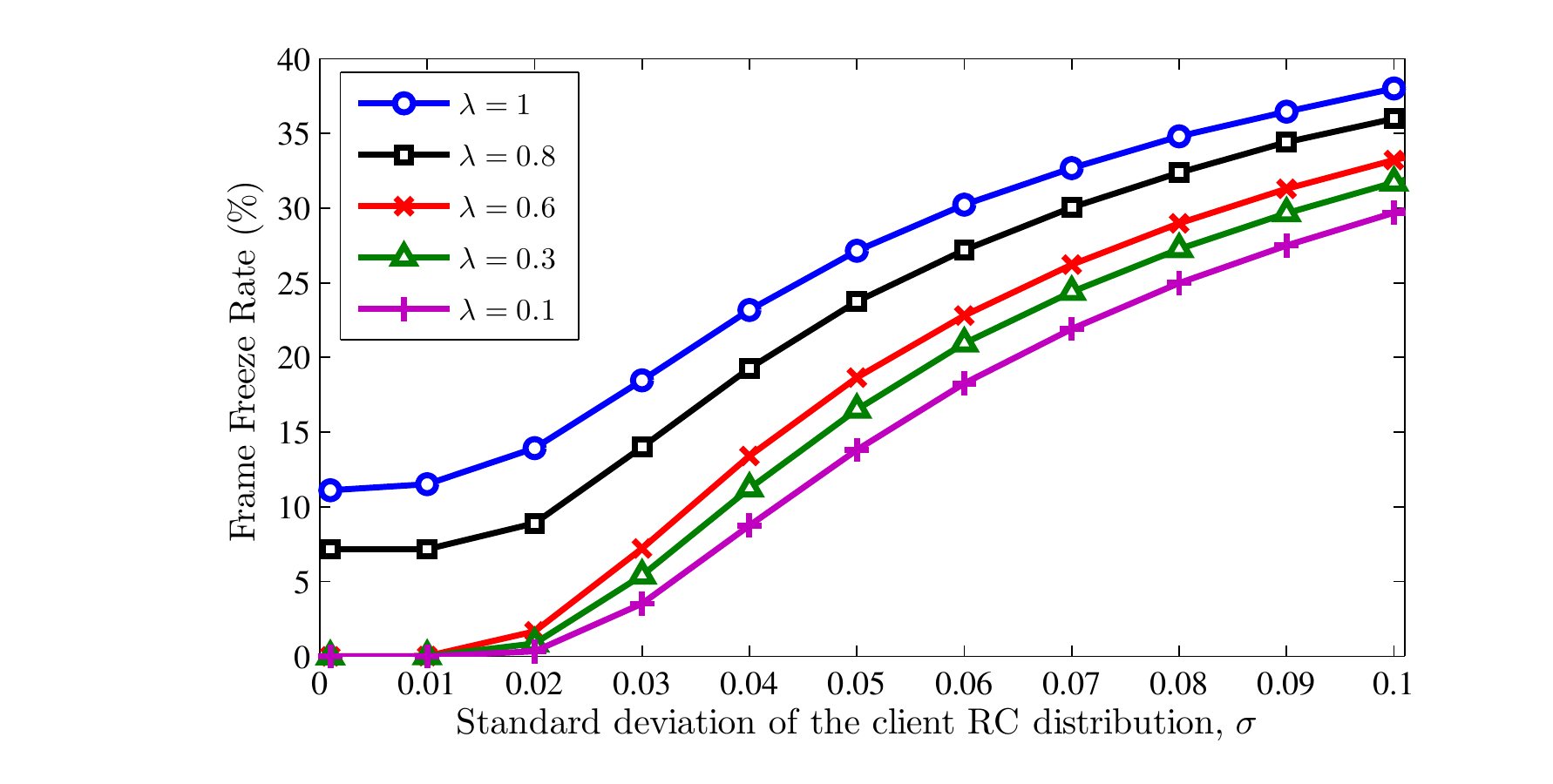}%
\caption{\csentence{FFR for dynamic optimization.} Frame freeze rate for the test client with time varying channel under dynamic optimization}\label{FFR_Dynamic}%
\end{figure}

\begin{figure}[h!]%
\includegraphics[width=12cm]{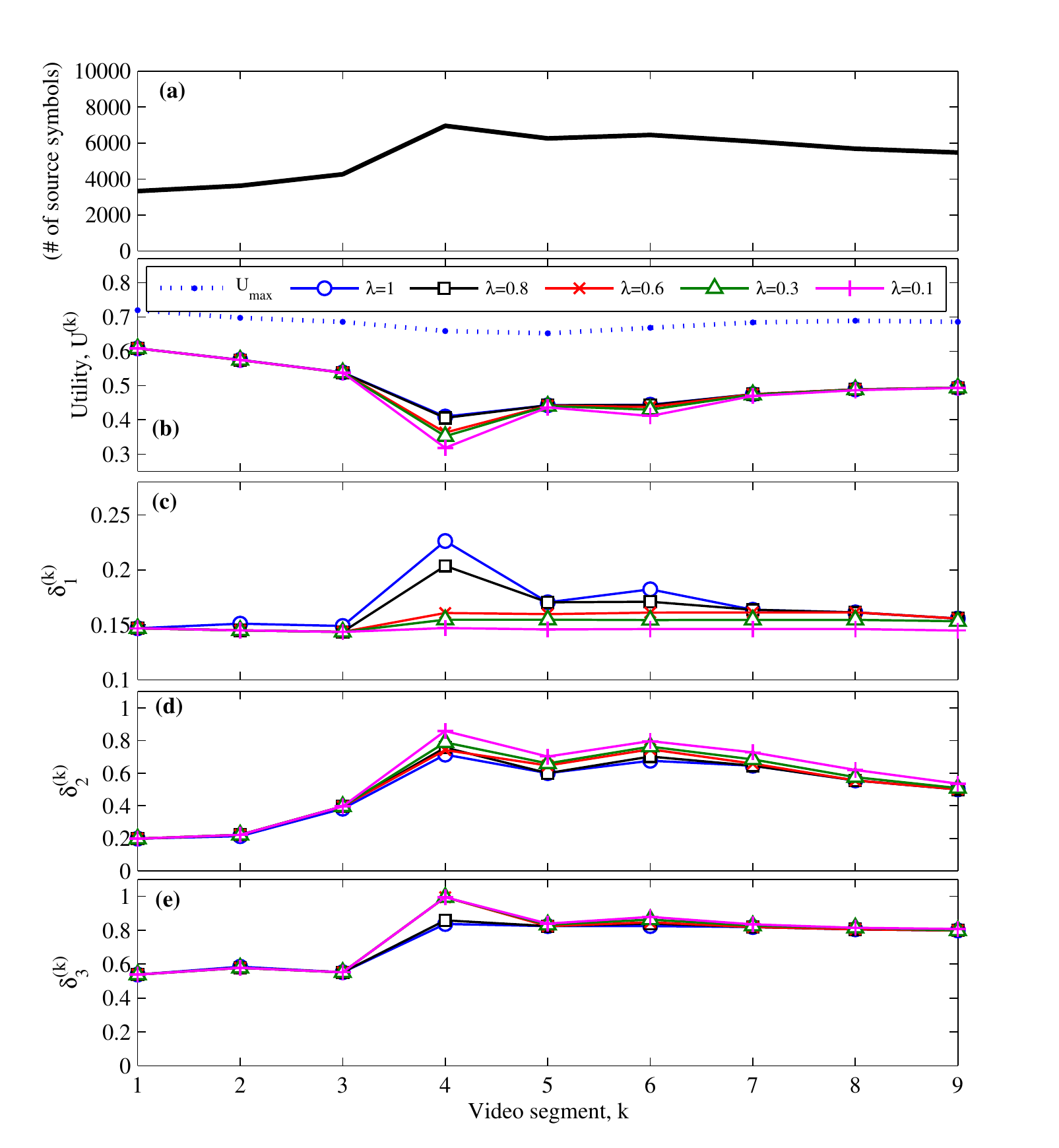}%
\caption{\csentence{Dynamic optimization Statistics.} Traces of (a) video segment size, (b) provided utility calculated using \eqref{equ:ServTot}, and (c-e) optimized MNRCs for the base-layer and the enhancement layers after solving Problem~3 with $N_{\text{max}}=11,000$.}%
\label{Fig_Dynamic}
\end{figure}


\section*{Tables}

\begin{table}[h!]\caption{Basic notations} \label{notations}
\begin{center}
     \begin{tabular}{l  p{12cm} }
\toprule
\textbf{Symbol} & \textbf{Definition}\\
\midrule
$L$ &  Total number of embedded layers in a video stream.\\
$S_l$ &  Number of source symbols per layer.\\
$M$ &  Total number of client classes.\\
$N_l$ &  Number of encoded symbols per layer.\\
$N_\text{max}$ &  Maximum amount of encoded symbols.\\  
$h_m$ &  Highest video layer that clients in class $m$ can potentially decode.\\
$0\leq\delta\leq 1$ &  Reception coefficient (RC) for a client.\\
$f_m(\delta) / F_m(\delta)$ &  RC probability distribution / cumulative distribution for class $m$.\\
$\pi_m$ &  Prior probability for class $m$ clients.\\
$P_{\text{out}}^l$ &  Outage probability constraint for layer $l$.\\
$P(S,N,\delta)$ &  Outage probability.\\

$\Omega$ &  Media server bandwidth (bit/s).\\
$B$ &  Size of the encoded symbols (bits).\\
$T_{\text{seg}}$ &  Duration of each video segment (s).\\
$R_l$ &  Cumulative source rate up to layer $l$ (bit/s).\\
$\mathcal{U}_m(R)$ &  Utility-rate function of class $m$.\\
$\alpha_{m,l}$ &  Incremental utility of layer $l$ for a client in class $m$.\\
$U_{m}$ &  Total utility for class $m$.\\
$\textrm{NMOS}$ &  Normalized mean opinion score.\\
\bottomrule
    \end{tabular} 
\end{center}
\end{table}

\begin{table}[h!]
\begin{center}
\caption{Specification of H.264/SVC coded video bitstreams.} \label{H264RD}
\resizebox{12cm}{!} {
\begin{tabular}{l c l c c c c }
  \toprule \toprule
 \multirow{2}{*}{Video} & \multirow{2}{*}{Layer} & {Resolution}   & {Frame Rate}   & {Bit Rate} & {Y-PSNR}  & {$S_l$}       \\ 
	{ } &  { } & {(pixels)}   & {(frames/s)}   & {(kbps)} & {(dB)}  & {(source symbols)}       \\ \midrule   
  \multirow{3}{*}{City} & {1} & {QCIF 176x144}    & {15}  & {104.3}  & {33.4}  & {261}       \\																	
  { } & {2} & {CIF { } 352x288}     & {30}  & {548.6}   & {33.5}  & {1111}       \\														
  { } & {3} & {4CIF 704x576}     & {60}  & {3226.2}  & {33.5}  & {6694}   \\   																
	\midrule
	\multirow{3}{*}{Ice} & {1} & {QCIF 176x144}    & {15}  & {84.6}  & {32.2}  & {212}       \\																	
  { } & {2} & {CIF { } 352x288}     & {30}  & {378.9}   & {34.9}  & {736}       \\														
  { } & {3} & {4CIF 704x576}     & {60}  & {2610.4}  & {38.6}  & {5579}   \\   																
	\midrule
	  \multirow{3}{*}{Crew} & {1} & {QCIF 176x144}    & {15}  & {150.8}  & {37.3}  & {377}       \\																	
  { } & {2} & {CIF { } 352x288}     & {30}  & {758.4}   & {37.1}  & {1519}       \\														
  { } & {3} & {4CIF 704x576}     & {60}  & {3560.4}  & {37.7}  & {7005}   \\   																

	\bottomrule
  \end{tabular}}
\end{center}
\end{table}

\begin{table}[!h]
\begin{center}
\caption{Performance of optimized allocation for the single-class scenario ($N_{\text{max}}=13,000$, $\alpha_l \triangleq \alpha_{1,l}, \forall l$).}
\label{Table_Results_Single}
\resizebox{12cm}{!} {
\begin{tabular}{ c c r r r r r r r r r r r r}
\toprule \toprule
{} & \multirow{3}{*}{\shortstack{Utility settings}}  & \multicolumn{4}{c}{City}  & \multicolumn{4}{c}{Ice}  & \multicolumn{4}{c}{Crew} \\ \cmidrule(r){3-6} \cmidrule(r){7-10} \cmidrule(r){11-14}
\multicolumn{1}{c}{\multirow{2}{*}{ {\shortstack{RC $\delta$ \\ Distribution}} }} & {}& \multicolumn{2}{c}{\multirow{1}{*}{Performance (\%) }} & \multicolumn{2}{c}{\multirow{1}{*}{Efficiency (\%) }} & \multicolumn{2}{c}{\multirow{1}{*}{Performance (\%) }} & \multicolumn{2}{c}{\multirow{1}{*}{Efficiency (\%) }} & \multicolumn{2}{c}{\multirow{1}{*}{Performance (\%) }} & \multicolumn{2}{c}{\multirow{1}{*}{Efficiency (\%) }}\\
 \cmidrule(r){3-4} \cmidrule(r){5-6} \cmidrule(r){7-8} \cmidrule(r){9-10} \cmidrule(r){11-12} \cmidrule(r){13-14}
 {} & {$[{\alpha_1}\quad {\alpha_2}\quad {\alpha_3}]$} & {$\eta^{^{CV}}\uparrow$} & {$\eta^{^{GD}}\uparrow$}  & {$\varepsilon^{^{CV}}$}  & {$\varepsilon^{^{GD}}$} & {$\eta^{^{CV}}\uparrow$} & {$\eta^{^{GD}}\uparrow$}  & {$\varepsilon^{^{CV}}$}  & {$\varepsilon^{^{GD}}$} & {$\eta^{^{CV}}\uparrow$} & {$\eta^{^{GD}}\uparrow$}  & {$\varepsilon^{^{CV}}$}  & {$\varepsilon^{^{GD}}$}\\[0pt] \midrule
\multicolumn{1}{c}{\multirow{4}{*}{$\Delta$-I}}  & {$[ {1}/{3}\quad {1}/{3}\quad {1}/{3}]$} &{114.40} &{114.40} &{100.00} &{100.00} &{71.22} &{71.22} &{100.00} &{100.00} &{113.82} &{113.82} &{100.00} &{100.00}  \\[0pt] 
 {} & {$[ {1}/{4}\quad {1}/{4}\quad {1}/{2}]$} &{81.74} &{81.74} &{100.00} &{100.00} &{52.40} &{52.40} &{100.00} &{100.00} &{75.89} &{75.89} &{100.00} &{100.00}  \\[0pt] 
 {} & {$[ {1}/{2}\quad {1}/{4}\quad {1}/{4}]$} &{144.48} &{144.48} &{100.00} &{100.00} &{87.18} &{87.18} &{100.00} &{100.00} &{152.44} &{152.44} &{100.00} &{100.00}  \\[0pt] 
 {} & {$[ {4}/{7}\quad {2}/{7}\quad {1}/{7}]$} &{176.61} &{176.61} &{100.00} &{100.00} &{105.33} &{105.33} &{100.00} &{100.00} &{188.40} &{188.40} &{100.00} &{100.00}  \\[0pt]\cmidrule{1-14} 

\multicolumn{1}{c}{\multirow{4}{*}{$\Delta$-II}}  &  {$[ {1}/{3}\quad {1}/{3}\quad {1}/{3}]$} &{15.83} &{20.53} &{96.10} &{100.00} &{14.43} &{22.55} &{93.37} &{100.00} &{17.96} &{25.36} &{94.10} &{100.00}  \\[0pt] 
 {} & {$[ {1}/{4}\quad {1}/{4}\quad {1}/{2}]$} &{11.58} &{16.79} &{95.54} &{100.00} &{ 8.37} &{16.44} &{93.06} &{100.00} &{17.91} &{21.34} &{97.18} &{100.00}  \\[0pt] 
 {} & {$[ {1}/{2}\quad {1}/{4}\quad {1}/{4}]$} &{22.49} &{27.82} &{95.83} &{100.00} &{20.06} &{26.68} &{94.77} &{100.00} &{24.52} &{32.87} &{93.71} &{100.00}  \\[0pt] 
 {} & {$[ {4}/{7}\quad {2}/{7}\quad {1}/{7}]$} &{27.96} &{31.48} &{97.32} &{100.00} &{26.12} &{31.10} &{96.20} &{100.00} &{28.79} &{38.39} &{93.06} &{100.00}  \\[0pt]\cmidrule{1-14} 

\multicolumn{1}{c}{\multirow{4}{*}{$\Delta$-III}} &{$[ {1}/{3}\quad {1}/{3}\quad {1}/{3}]$} &{281.10} &{318.43} &{90.42} &{99.27} &{379.38} &{395.35} &{96.77} &{99.99} &{209.62} &{210.07} &{98.75} &{98.89}  \\[0pt] 
 {} & {$[ {1}/{4}\quad {1}/{4}\quad {1}/{2}]$} &{167.02} &{213.91} &{84.47} &{99.30} &{238.25} &{293.25} &{86.01} &{99.99} &{113.42} &{154.63} &{83.81} &{99.99}  \\[0pt] 
 {} & {$[ {1}/{2}\quad {1}/{4}\quad {1}/{4}]$} &{358.66} &{395.93} &{92.49} &{100.00} &{434.55} &{454.58} &{96.39} &{100.00} &{326.65} &{336.00} &{97.85} &{100.00}  \\[0pt] 
 {} & {$[ {4}/{7}\quad {2}/{7}\quad {1}/{7}]$} &{424.19} &{466.78} &{92.49} &{100.00} &{514.46} &{530.42} &{97.47} &{100.00} &{387.60} &{398.29} &{97.85} &{100.00}  \\[0pt]\cmidrule{1-14} 

\multicolumn{1}{c}{\multirow{4}{*}{$\Delta$-IV}} &  {$[ {1}/{3}\quad {1}/{3}\quad {1}/{3}]$} &{38.68} &{54.54} &{89.73} &{100.00} &{32.92} &{40.35} &{90.02} &{95.04} &{32.22} &{34.00} &{95.58} &{96.87}  \\[0pt] 
 {} & {$[ {1}/{4}\quad {1}/{4}\quad {1}/{2}]$} &{27.40} &{27.83} &{91.47} &{91.78} &{22.02} &{31.10} &{93.07} &{100.00} &{23.06} &{24.59} &{97.07} &{98.28}  \\[0pt] 
 {} & {$[ {1}/{2}\quad {1}/{4}\quad {1}/{4}]$} &{61.61} &{72.15} &{93.88} &{100.00} &{50.69} &{53.27} &{95.22} &{96.85} &{41.34} &{59.93} &{88.37} &{99.99}  \\[0pt] 
 {} & {$[ {4}/{7}\quad {2}/{7}\quad {1}/{7}]$} &{78.11} &{83.92} &{96.84} &{100.00} &{61.30} &{70.15} &{94.80} &{100.00} &{57.03} &{72.07} &{91.01} &{99.72}  \\[0pt]\cmidrule{1-14} 

 \multicolumn{2}{c}{\textbf{     Average}} &  \textbf{126.99} &\textbf{140.46} &\textbf{94.79} &\textbf{99.40} &\textbf{132.42} &\textbf{142.59} &\textbf{95.45} &\textbf{99.49} &\textbf{113.17} &\textbf{121.13} &\textbf{95.52} &\textbf{99.61}  \\[0pt]
\bottomrule
\end{tabular}
}
\end{center}
\end{table}

\begin{table}[!h]
\begin{center}
\caption{Simulation results for $M=2$ classes and $L=3$ layers ($\pi_2=1-\pi_1$).}
\label{Table_Results}
\resizebox{12cm}{!} {
\begin{tabular}{ c c r r r r r r r r r r r r}
\toprule \toprule
\multicolumn{1}{c}{\multirow{4}{*}{ $N_{\text{max}}$ }} & \multicolumn{1}{c}{\multirow{4}{*}{ $\pi_1$ }}  & \multicolumn{4}{c}{City}  & \multicolumn{4}{c}{Ice}  & \multicolumn{4}{c}{Crew} \\ \cmidrule(r){3-6} \cmidrule(r){7-10} \cmidrule(r){11-14}
 {} & {} & \multicolumn{2}{c}{\multirow{1}{*}{Performance (\%) }}  & \multicolumn{2}{c}{\multirow{1}{*}{Efficiency (\%) }} & \multicolumn{2}{c}{\multirow{1}{*}{Performance (\%) }}  & \multicolumn{2}{c}{\multirow{1}{*}{Efficiency (\%) }} & \multicolumn{2}{c}{\multirow{1}{*}{Performance (\%) }}  & \multicolumn{2}{c}{\multirow{1}{*}{Efficiency (\%) }}\\
\cmidrule(r){3-4} \cmidrule(r){5-6} \cmidrule(r){7-8} \cmidrule(r){9-10} \cmidrule(r){11-12} \cmidrule(r){13-14}
 {} & {} & {$\eta^{^{CV}}\uparrow$} & {$\eta^{^{GD}}\uparrow$}  & {$\varepsilon^{^{CV}}$}  & {$\varepsilon^{^{GD}}$}  & {$\eta^{^{CV}}\uparrow$} & {$\eta^{^{GD}}\uparrow$}  & {$\varepsilon^{^{CV}}$}  & {$\varepsilon^{^{GD}}$}  & {$\eta^{^{CV}}\uparrow$} & {$\eta^{^{GD}}\uparrow$}  & {$\varepsilon^{^{CV}}$}  & {$\varepsilon^{^{GD}}$}\\[0pt] \midrule

\multicolumn{1}{c}{\multirow{5}{*}{$10,000$}} & {0.1} &{139.75} &{185.02} &{88.71} &{99.41} &{94.55} &{95.79} &{98.36} &{99.25} &{729.00} &{840.57} &{93.21} &{97.99} \\[0pt] 
 {} & {0.3} &{148.06} &{180.00} &{92.00} &{99.71} &{91.98} &{93.47} &{98.59} &{99.59} &{694.42} &{774.50} &{96.11} &{99.23} \\[0pt] 
 {} & {0.5} &{164.39} &{193.34} &{93.25} &{99.99} &{97.00} &{100.11} &{97.52} &{99.45} &{711.10} &{793.00} &{96.85} &{99.77} \\[0pt] 
 {} & {0.7} &{187.93} &{222.15} &{92.66} &{100.00} &{116.24} &{118.15} &{98.30} &{99.34} &{754.36} &{842.38} &{96.01} &{99.09} \\[0pt] 
 {} & {0.9} &{230.60} &{284.35} &{91.00} &{100.00} &{158.42} &{159.80} &{99.24} &{99.98} &{868.80} &{1001.10} &{92.76} &{97.69} \\[0pt]\cmidrule{1-14}

\multicolumn{1}{c}{\multirow{5}{*}{$15,000$}} & {0.1}  &{88.06} &{95.91} &{95.54} &{99.70} &{86.72} &{94.33} &{95.20} &{100.00} &{99.03} &{103.91} &{97.26} &{99.98} \\[0pt] 
 {} & {0.3} &{83.34} &{90.59} &{95.81} &{99.91} &{81.91} &{87.50} &{96.32} &{100.00} &{90.59} &{94.45} &{97.59} &{99.84} \\[0pt] 
 {} & {0.5} &{90.77} &{95.43} &{97.05} &{99.94} &{86.43} &{90.20} &{97.47} &{100.00} &{93.05} &{96.36} &{97.77} &{99.62} \\[0pt] 
 {} & {0.7} &{108.87} &{111.70} &{98.01} &{99.84} &{100.93} &{103.02} &{98.56} &{99.97} &{106.31} &{109.35} &{98.30} &{99.88} \\[0pt] 
 {} & {0.9} &{152.94} &{154.14} &{99.28} &{99.89} &{137.00} &{138.04} &{99.37} &{99.92} &{134.99} &{138.98} &{98.33} &{99.98} \\[0pt]\cmidrule{1-14}

\multicolumn{1}{c}{\multirow{5}{*}{$19,000$}}  & {0.1} &{78.32} &{86.74} &{94.92} &{100.00} &{47.77} &{50.33} &{98.24} &{100.00} &{102.87} &{110.00} &{95.79} &{99.95} \\[0pt] 
 {} & {0.3} &{74.52} &{80.94} &{95.86} &{100.00} &{49.81} &{51.74} &{98.68} &{100.00} &{93.00} &{97.82} &{96.92} &{99.98} \\[0pt] 
 {} & {0.5} &{79.53} &{83.96} &{97.03} &{100.00} &{54.77} &{56.13} &{99.07} &{100.00} &{95.44} &{98.64} &{98.00} &{99.95} \\[0pt] 
 {} & {0.7} &{93.92} &{96.50} &{98.26} &{99.98} &{63.90} &{64.76} &{99.40} &{100.00} &{108.67} &{110.06} &{98.62} &{99.95} \\[0pt] 
 {} & {0.9} &{128.10} &{129.14} &{99.33} &{99.89} &{80.97} &{81.63} &{99.61} &{99.98} &{142.63} &{144.29} &{99.35} &{99.97} \\[0pt]\cmidrule{1-14} 

\multicolumn{2}{c}{\textbf{     Average}}  &\textbf{170.55} &\textbf{189.21} &\textbf{96.53} &\textbf{99.92} &\textbf{125.46} &\textbf{126.49} &\textbf{99.41} &\textbf{99.96} &\textbf{321.61} &\textbf{357.09} &\textbf{96.87} &\textbf{99.52} \\[0pt]
\bottomrule
\end{tabular}
}
\end{center}
\end{table}

\begin{table}[!h]
\begin{center}
\caption{$\overline{\mathcal{D}}:$ Percentage of clients that experience outage in decoding the base layer. $\mathcal{Z}:$ Percentage of clients that experience outage  at least once over the 9 video segments.} \label{Dynamic}
\resizebox{12cm}{!} {
\begin{tabular}{r r r r r r r r r r r r r r }
  \toprule \toprule
  \multirow{2}{*}{$N_{\text{max}}  $} &  \multicolumn{6}{c}{\multirow{1}{*}{$\overline{\mathcal{D}}$ (\%)}  } & { } & \multicolumn{6}{c}{\multirow{1}{*}{$\mathcal{Z}$ (\%)}  }   \\ \cmidrule(r){2-7}  \cmidrule(r){9-14}
	 {}   & {EEP} & {$\lambda=1$}   & {$\lambda=0.8$} & {$\lambda=0.6$} & {$\lambda=0.3$} & {$\lambda=0.1$} & { } & {EEP} & {$\lambda=1$}   & {$\lambda=0.8$} & {$\lambda=0.6$} & {$\lambda=0.3$} & {$\lambda=0.1$}       \\ \midrule
  {9000} & {  3.38} & {  4.40} & {  3.62} & {  2.51} & {  1.89} & {  1.89} & { } & { 58.20} & { 38.81} & { 34.47} & { 25.00} & { 19.40} & { 19.40} \\ 
 {10000} & {  2.16} & {  3.50} & {  2.75} & {  1.17} & {  0.53} & {  0.52} & { } & { 46.33} & { 27.29} & { 25.19} & { 12.43} & {  6.71} & {  6.71} \\ 
 {11000} & {  1.87} & {  1.70} & {  1.03} & {  0.22} & {  0.12} & {  0.04} & { } & { 42.02} & { 16.03} & { 11.73} & {  4.39} & {  3.62} & {  2.88} \\ 
 {12000} & {  1.78} & {  0.74} & {  0.35} & {  0.25} & {  0.11} & {  0.00} & { } & { 39.61} & {  7.16} & {  5.19} & {  4.31} & {  3.04} & {  2.14} \\ 
 {13000} & {  1.71} & {  0.41} & {  0.17} & {  0.08} & {  0.01} & {  0.00} & { } & { 37.57} & {  4.89} & {  3.21} & {  2.35} & {  1.71} & {  1.67} \\ 
 {14000} & {  1.68} & {  0.30} & {  0.15} & {  0.06} & {  0.00} & {  0.00} & { } & { 35.75} & {  4.08} & {  2.68} & {  1.86} & {  1.34} & {  1.34} \\ 
 {15000} & {  1.71} & {  0.36} & {  0.15} & {  0.07} & {  0.01} & {  0.00} & { } & { 34.15} & {  4.06} & {  2.37} & {  1.67} & {  1.09} & {  1.06} \\ 
 {16000} & {  1.80} & {  0.37} & {  0.17} & {  0.09} & {  0.01} & {  0.00} & { } & { 32.73} & {  3.83} & {  2.37} & {  1.67} & {  0.93} & {  0.87} \\ 
\bottomrule
  \end{tabular}}
\end{center}
\end{table}


%

\end{backmatter}
\end{document}